\documentclass[acmlarge]{acmart}

\usepackage{amsmath,amssymb,amsfonts}
\usepackage{algorithmic}
\usepackage{graphicx}
\usepackage{textcomp}
\usepackage{xcolor}
\usepackage{booktabs}
\usepackage{subcaption}
\usepackage{enumitem}
\usepackage{pifont}
\usepackage{epstopdf}
\usepackage[utf8]{inputenc}
\usepackage{tablefootnote}
\usepackage{url}
\usepackage{multirow}

\usepackage{dblfloatfix}

\def\BibTeX{{\rm B\kern-.05em{\sc i\kern-.025em b}\kern-.08em
    T\kern-.1667em\lower.7ex\hbox{E}\kern-.125emX}}

\setcopyright{acmlicensed}
\copyrightyear{2025}
\acmYear{2025}
\acmDOI{XXXXXXX.XXXXXXX}

\newcommand{\todo}[1]{\textcolor{red}{#1}}

\newcommand{\rev}[1]{\textcolor{black}{#1}}

\newcommand{\bh}[1]{\vspace{3pt}\noindent \textbf{#1} }

\newcommand{\highlight}[1]{\textbf{\textcolor{red!60!black}{#1}}}

\begin{document}
% \title{MFIT : \underline{M}ulti\underline{FI}delity \underline{T}hermal Modeling for 2.5D and 3D Chiplet-based Systems \todo{on Interposers}}
\title{MFIT : \underline{M}ulti-\underline{FI}delity \underline{T}hermal Modeling for 2.5D and 3D Multi-Chiplet Architectures}
\author{Lukas Pfromm}
\authornote{Both authors contributed equally to this research.}
\email{pfromm@wisc.edu}
\author{Alish Kanani}
\authornotemark[1]
\email{ahkanani@wisc.edu}
\affiliation{%
  \institution{University of Wisconsin–Madison}
  \city{Madison}
  \state{Wisconsin}
  \country{USA}
}

\author{Harsh Sharma}
\affiliation{%
  \institution{Washington State University}
  \city{Pullman}
  \state{Washington}
  \country{USA}
}

\author{Parth Solanki}
\author{Eric Tervo}
\affiliation{%
  \institution{University of Wisconsin-Madison}
  \city{Madison}
  \state{Wisconsin}
  \country{USA}
}

\author{Jaehyun Park}
\affiliation{%
  \institution{University of Ulsan}
  \city{Ulsan}
  \country{Republic of Korea}
}

\author{Janardhan Rao Doppa}
\author{Partha Pratim Pande}
\affiliation{%
  \institution{Washington State University}
  \city{Pullman}
  \state{Washington}
  \country{USA}
}

\author{Umit Y. Ogras}
\affiliation{%
  \institution{University of Wisconsin–Madison}
  \city{Madison}
  \state{Wisconsin}
  \country{USA}
}

% }

% \thanks{\IEEEauthorrefmark{1} Equal contribution
% \textit{Corresponding Author: Lukas Pfromm (pfromm@wisc.edu)}
% L. Pfromm, A. Kanani, E. Tervo and U. Y. Ogras are with the Department of Electrical and Computer Engineering, University of Wisconsin–Madison, WI, USA.
% H. Sharma, J. R. Doppa and P. P. Pande are with the School of Electrical Engineering and Computer Science, Washington State University, WA, USA.
% P. Solanki and E. Tervo are with the Department of Mechanical Engineering, University of Wisconsin-Madison, WI, USA.
% J. Park is with Department of Electrical, Electronic and Computer Engineering, University of Ulsan, Republic of Korea.
% }

% \thanks{\IEEEauthorrefmark{5} equal contribution}

\date{}
% \textbf{}

\begin{abstract}
Rapidly evolving artificial intelligence and machine learning applications require ever-increasing computational capabilities, while monolithic 2D design technologies approach their limits. 
2.5D/3D heterogeneous integration of smaller chiplets using advanced packaging has emerged as a promising paradigm for addressing this limit and meeting performance demands.
These approaches offer a significant cost reduction and higher manufacturing yield than monolithic 2D integrated circuits. 
However, the compact arrangement and high compute density of these systems exacerbate thermal management challenges, potentially compromising performance. 
Addressing these thermal modeling challenges is critical, especially as system sizes grow and different design stages require varying levels of accuracy and speed. Since no single thermal modeling technique meets all these needs, this paper introduces MFIT, a range of \textit{multi-fidelity thermal models} that effectively balance accuracy and speed. These multi-fidelity models can enable efficient design space exploration and runtime thermal management. 
Our extensive testing on systems with 16, 36, and 64 2.5D integrated chiplets and 16$\times$3 3D integrated chiplets demonstrates that these models can reduce execution times from days to mere seconds and milliseconds with negligible loss in accuracy.

\end{abstract}

\maketitle

%\vspace{-2mm}
\section{Introduction} \label{sec:introduction}
% High data and processing requirements require new processing approaches. Traditional monolithic chips cannot meet this requirement due to cost and yield. 
Massive data from different modalities, including text, images, video, and speech, are continuously produced by various sensors. At the same time, increasingly complex artificial intelligence (AI) and machine learning (ML) algorithms process this data to enable new applications that were previously impractical. This trend dictates the design of large-scale chips with high memory and compute capabilities, offering a high degree of parallelism~\cite{src_report, IRTS2015}. Traditional 2D chip design and packaging technologies cannot sustain this need due to the low yield of large monolithic planar chips and the corresponding increase in fabrication cost~\cite{Next_dtco}.
Therefore, new design approaches are required to meet the increasing demand for computing power and memory capacity~\cite{src_report}.

\begin{figure}[t]
    \centering
    %\vspace{3mm}
    
    \begin{subfigure}{0.48\columnwidth}
        \centering
        \includegraphics[width=\columnwidth]{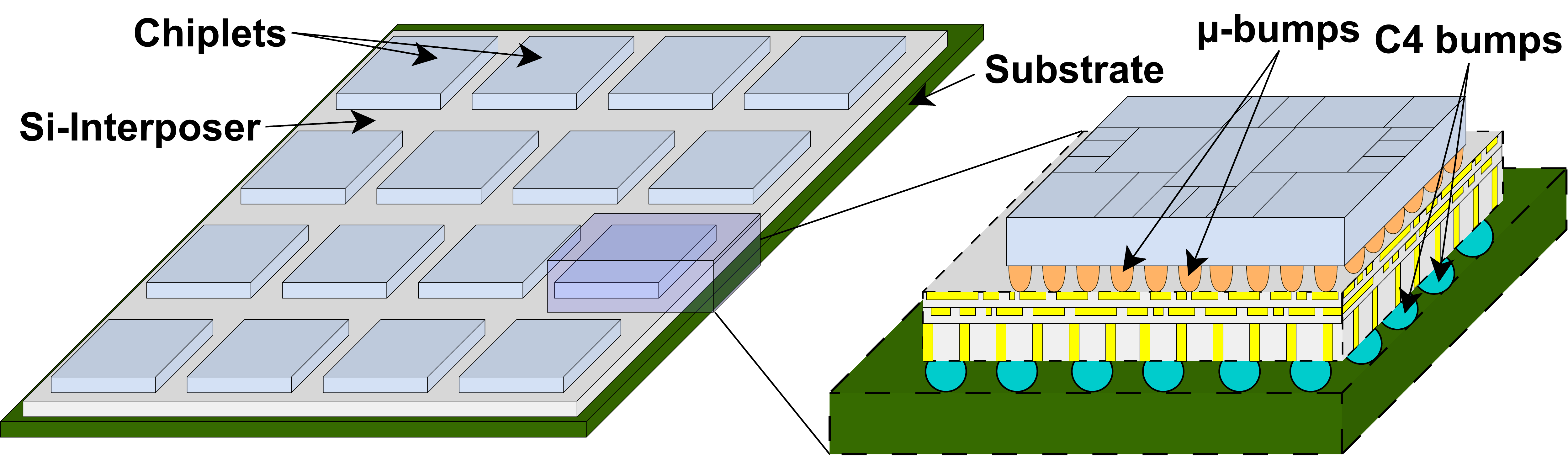}
        % \Description{Illustration of a 16-chiplet system}
        % \vskip -5pt
        \caption{A 16 - 2.5D integrated chiplet-based system. 
        The magnified view on the right shows a detailed structure of a single chiplet.}
        \label{fig:simplified_4x4_top}
    \end{subfigure}
    \hfill
    \begin{subfigure}{0.48\columnwidth}
        \centering
        \includegraphics[width=\columnwidth]{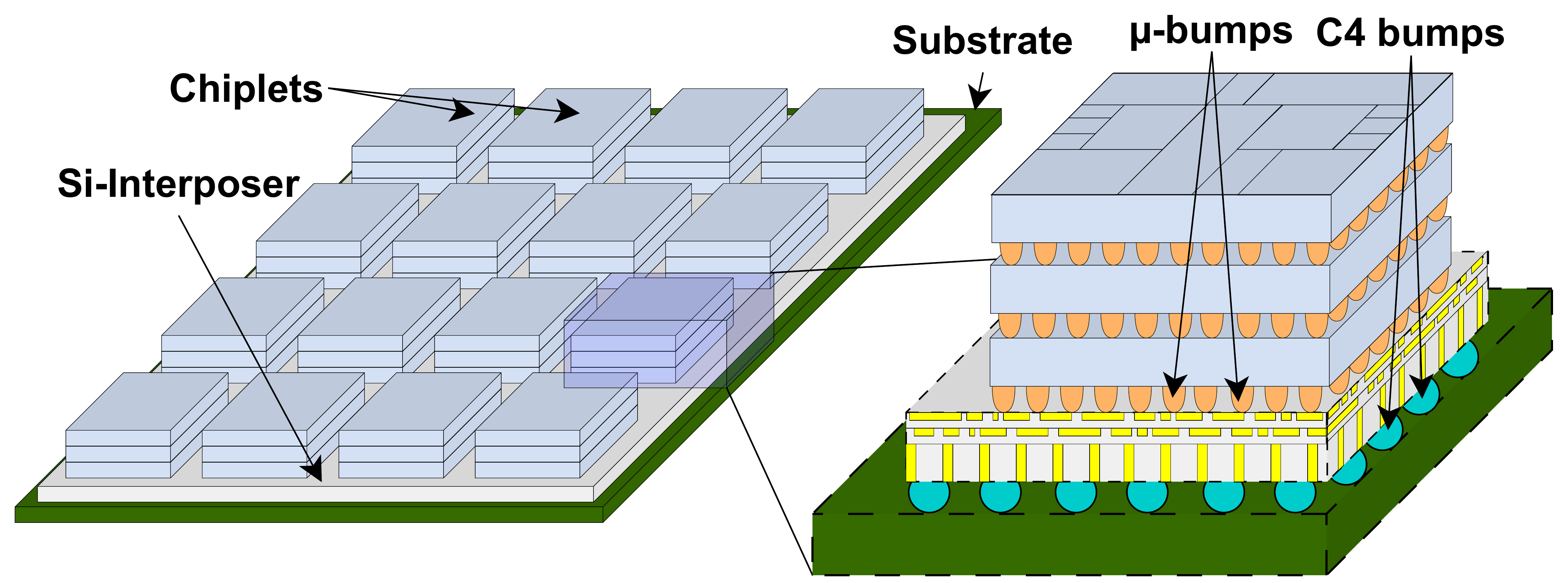}
        % \Description{Illustration of a 16-chiplet system}
        % \vskip -5pt
        \caption{A 16$\times$3 - 3D integrated chiplet-based system. The magnified view on the right shows a detailed structure of three chiplets stacked.}
        \label{fig:simplified_4x4_bottom}
    \end{subfigure}

    \vskip -10pt
    
    \caption{2.5D/3D integrated chiplet systems considered in this work, showing the chiplets, interposer, and part of the substrate.}
    \label{fig:stacked_simplified_4x4}
\end{figure}

% Chiplets are cheaper to manufacture, scale more efficiently, and can be integrated heterogeneously
2.5D and 3D chiplet-based architectures have emerged as promising alternatives to traditional monolithic 2D chips due to their lower fabrication costs~\cite{simba, sharma2023florets, krishnan2022biglittle}.
Compared to conventional monolithic systems, chiplet-based systems integrate multiple small pre-fabricated chips (chiplets) on a silicon interposer, which facilitates data exchange, as illustrated in \rev{Fig.}~\ref{fig:simplified_4x4_top}.
3D packaged systems expand on this approach by stacking multiple chiplets vertically and connecting them with vertical vias, creating a more compact system as illustrated in \rev{Fig.}~\ref{fig:simplified_4x4_bottom}.
The smaller size of these chiplets enables a higher yield and lower overall manufacturing cost than traditional monolithic dies~\cite{stow2016costAnalysis}.
Additionally, this modular approach facilitates scaling the system sizes and enables heterogeneous integration of different chiplet types, e.g., memory, processing, and processing-in-memory chiplets.
Hence, emerging 2.5D and 3D architectures enable a new cost-effective avenue for compact scale-out implementations of various emerging compute- and data-intensive applications, including AI/ML. Indeed, these advantages have led to industrial adoption by companies including Intel~\cite{intel2023agilex, intel2006dieStacking}, AMD~\cite{amd2021chipletTechnology, kite, agarwal20223d}, and NVIDIA~\cite{simba}.

% This section on the importance of thermal modeling
% Umit: This paragraph is good. We can remove color coding and polish it.
Thermal bottlenecks have long been a significant barrier to increasing the performance of computing systems. 2.5D and 3D integrated systems exacerbate this barrier due to their dense integration and unique physical structure~\cite{park2024thermal}. 
\rev{Unlike monolithic chips, where heat spreads uniformly across a continuous die, 2.5D chiplet-based systems conduct heat laterally through the interposer and vertically through the heat spreader. 
Although the interposer, often built with materials of relatively high thermal conductivity, can help alleviate localized thermal hotspots, chiplet-based designs introduce new sources of thermal stress. 
In particular, a single package can accommodate more active silicon by integrating multiple smaller chiplets, thereby increasing the total compute throughput and system power. If the total power consumption increases faster than the area, the package-level power density also becomes larger than a monolithic chip. 
Furthermore, while the interposer aids in thermal spreading, it also hosts numerous communication links whose resistive Joule heating contributes to additional thermal challenge~\cite{ma2024electrical}.}
% In contrast to a monolithic chip, where heat is spread directly across the die, a 2.5D chiplet-based system conducts heat between different chiplets through the interposer and heat spreader. 
Likewise, heat also flows vertically between adjacent stacked chiplets in a 3D chiplet-based system.
% Moreover, an adjacent stacked chiplet is not always a cooling path but can be a heat source, increasing thermal density.

% Unlike a monolithic chip, where heat is spread directly across the die, a chiplet-based system conducts heat between different chiplets through the interposer and heat spreader. 
These factors introduce unique challenges for effective thermal management in these systems. Traditional design flows and physical floorplanning focus on reducing wire lengths to meet timing constraints and minimizing area to reduce fabrication costs. However, these objectives could also lead to thermal crosstalk, thermal hotspots, and compromise performance.
% New paragraph
Chiplet-based systems introduce additional design parameters such as inter-chiplet link length, spacing, chiplet placement, sizing, inter-layer communication, and design partitioning. Tuning traditional and chiplet-based design parameters while maintaining thermal stability is critical to ensure a thermally-efficient design.
% Old paragraph
% Hence, balancing traditional design parameters with thermal stability by tuning chiplet-specific design parameters (e.g., inter-chiplet link length, spacing, chiplet placement, sizing, inter-layer communication, and design partitioning) is critical to ensure a thermally-efficient design.

% \begin{figure}[t]
% \centering
%     \centering
%     \includegraphics[width=\columnwidth]{figures/ubump_experiments/ubump_2chiplet_contour.png}
%     \caption{Two chiplets on a silicon interposer. One chiplet is actively generating heat, while the other is inactive. This figure illustrates how heat generated by the active chiplet flows through the silicon interposer and heat spreader.}
%     \label{fig:2chiplet_contour}
% \end{figure}

% This is the figure showing the comparison of each of the four models in speed, accuracy, and use case
\begin{figure*}[ht]
\centering
    \centering
    % \vskip -10pt
    \includegraphics[width=0.98\linewidth]{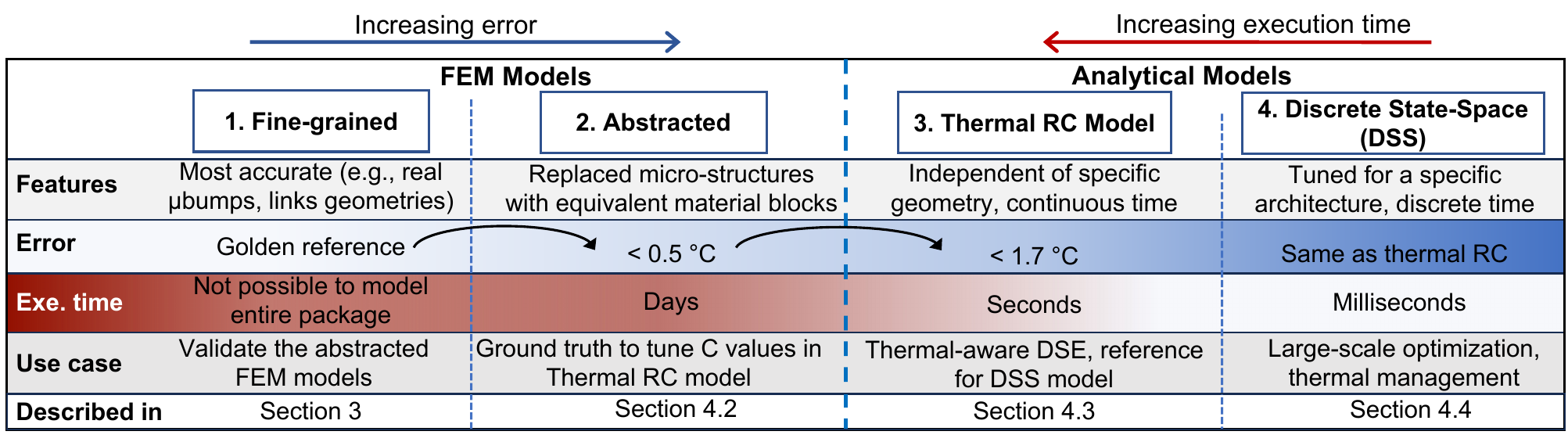}
    % \Description{...}
    % \vskip -10pt
    \caption{ Summary of the multi-fidelity thermal models. (1) Fine-grained FEM models capture precise geometry but are too complex to simulate the entire chiplet-based system. 
    (2) Abstracted FEM models are derived from the fine-grained model to simulate large-scale systems with negligible impact on accuracy ($<\mathrm{0.5}^{\circ}$C, which is 0.5--1\% around the temperatures of interest). 
    (3) Since abstract FEM models are still too slow for DSE, they are used to tune thermal RC circuit models, which introduce less than 1.7$^{\circ}$C (1--3.5\% around the temperatures of interest) error).
    (4) Further abstraction reduces the execution time to milliseconds using DSS models developed for specific system configurations, enabling runtime thermal management.
    } 
    \label{fig:overview}
    \vskip -10pt
\end{figure*}  

The semiconductor chip design cycle spans multiple phases: system specification, architecture exploration, logic design, physical design and validation, fabrication, and post-silicon optimization/validation. 
Each phase has a unique set of design constraints and requirements. For example, the lack of a test chip during the pre-silicon phases requires simulation and analytical models. 
Finite Element Method (FEM) simulations offer the most accurate approach for pre-silicon thermal analysis~\cite{sultan2019thermalSimulatorSurvey}. 
They can serve as a reference and enable heat flow studies to guide the design process. 
However, they are too slow for practical architecture and design space exploration (DSE), as illustrated in \rev{Fig.}~\ref{fig:overview}. 
Modeling the package as a thermal RC (resistive-capacitive) network can significantly accelerate simulations with acceptable accuracy loss~\cite{skadron_temperatureAware,3D_ICE}. 
Since each node in the thermal circuit corresponds to a specific location in the package, thermal RC models solve discretized versions of the FEM models \textit{in space}. Hence, they enable thermally-aware DSE and optimization with a finite number of discrete hotspot nodes.
However, the thermal resistance/capacitance values and the circuit topology must accurately reflect the chip geometry and material properties for reliable results. 
%Hence, the thermal RC models must be verified for accuracy against FEM or hardware measurements. 
Since the thermal RC models solve continuous-time ordinary differential equations (ODEs), they have execution times in the order of seconds to minutes. Therefore, they cannot be used for runtime optimization tasks such as dynamic thermal and power management (DTPM). 
With a given sampling period, one can discretize them \textit{in the time domain}. The resulting discrete state-space (DSS) models significantly reduce runtime \textit{at the cost of further abstracting the model from the physical package}. Consequently, they apply only to the specific configurations for which they are developed.

There is a strong need for tools to accurately analyze the thermal behavior of 2.5D and 3D integrated systems and guide their design process. However, no single modeling technique can alone address the needs of all design phases.
To address this critical gap, this paper proposes MFIT, a \textit{multi-fidelity thermal modeling framework} that synergistically exploits the strengths of each class of models (FEM, thermal RC, and DSS). We use this framework to produce a set of thermal models that can guide the entire design cycle, unlike a point solution that can serve a specific portion of the design process. 
\textit{The elements of this set not only cover complementary parts of the design cycle but support each other and produce consistent results.} 
We first develop a fine-grained FEM model of the target package as a reference. Since it is slow and computationally expensive, we next judiciously design an abstracted version of this fine-grained FEM model to simulate an entire package in days while maintaining accuracy. 
%For example, replacing individual $\mu$-bumps and links by single blocks enable 3--4-fold speed-up with negligible accuracy impact, as shown in Figure~\ref{fig:overview}. 
%The abstracted FEM models can model an entire package and produce an accurate reference design. However, they are still impractical in evaluating future large-scale systems with tens and hundreds of chiplets. 
To enable fast DSE, MFIT also incorporates thermal RC circuit models verified against the reference FEM models. 
Our thermal RC models run in the order of seconds while leading to less than 1.7$^{\circ}$C error when the package temperature is around 100$^{\circ}$C, as summarized in \rev{Fig.}~\ref{fig:overview}. 
Hence, they can be used for pre-silicon architectural optimization, such as mapping the workloads to chiplets, network-on-interposer design, and chiplet placement for 2.5D and 3D stacked systems.
Finally, MFIT derives a final class of models by discretizing the thermal RC models, enabling runtime thermal management and large-scale DSE in the order of milliseconds. 
% The resulting discrete state-space models produce nearly identical results to the RC models from which they are derived. 
Since these discrete models are generated for a specific sampling period and configuration, they are regenerated from the RC model if the target configuration changes \textit{automatically in a few milliseconds.}
In summary, we obtain a set of multi-fidelity thermal models that guide and complement each other to cover all design phases. 

The key contributions of this work are as follows:
%\vspace{-2mm}
\begin {itemize}[leftmargin=*]
    \item A novel thermal modeling approach that systematically abstracts fine-grained FEM models to produce abstract FEM, thermal RC, and DSS models to achieve varying speed and accuracy trade-offs,

    \item A family of \textit{open-source multi-fidelity thermal models} that span a broad accuracy (reference to 1.7$^{\circ}$C) and speed (days to milliseconds) range,
    
    \item Extensive evaluations with 16, 36, and 64 - 2.5D and 16$\times$3 - 3D integrated chiplets systems running AI/ML workloads to demonstrate the accuracy and speed-up benefits of our multi-fidelity thermal models,

    \item Open-sourced code for thermal RC and DSS models at \href{https://github.com/AlishKanani/MFIT}{github.com/AlishKanani/MFIT}. 
    The FEM models created for this work are included as well.

\end {itemize}

The remainder of the paper is organized as follows. Section~\ref{sec:related_work} and Section~\ref{sec:fem_overview} discuss related work and background on FEM. Section~\ref{sec:modeling_approach} presents the proposed multi-fidelity thermal modeling framework. Finally, Section~\ref{sec:experimental_results} presents the experimental evaluation, and Section~\ref{sec:conclusion} concludes the paper.

%\vspace{-2mm}
\section{Related Work} \label{sec:related_work}

2.5D and 3D integration-based systems are becoming mainstream due to higher performance and lower manufacturing costs than monolithic chips.
Both domain-specific and general-purpose 2.5D and 3D architectures have been explored to date~\cite{kite, krishnan2021siam, sharma2022swap, sharma2023florets, simba, eckert2014thermal, agarwal20223d}. 
SIMBA is one of the first prototype multi-chip modules with 36 chiplets designed for inference with deep models~\cite{simba}. 
% SWAP is an application-specific 2.5D accelerator that leverages the expected traffic patterns to optimize the interconnect architecture for performance and energy efficiency~\cite{sharma2022swap}. 
Similarly, Floret is a data center-scale architecture for accelerating convolutional neural network (CNN) inference tasks by exploiting dataflow patterns~\cite{sharma2023florets}. 
% It proposes a space-filling curve (SFC)-based network-on-interposer (NoI) to connect chiplets on a contiguous path. 
% {\em Existing chiplet architectures focus on performance and energy efficiency without evaluating thermal aspects extensively.}
Loi et al.~\cite{loi2006thermally} analyze the performance benefits of vertically integrated (3D) processor-memory hierarchy
under thermal constraints. 
Similarly, Eckert et al.~\cite{eckert2014thermal} consider processing-in-memory (PIM) architectures implemented using 3D die stacking. They study the thermal constraints across different processor organizations and cooling solutions to identify viable solutions.
Our proposed open-source thermal models catalyze similar thermal analysis and optimization studies for 2.5D and 3D integrated systems.
% Multiple works have been previously proposed to study the thermal profile during the chip design. However, these approaches are constrained by their singular focus on specific design phases.
% % by the accuracy and time taken to undergo experimentation. This limits the possibility of an effective joint-performance-thermal trade-off in the design process. 

The most accurate and direct thermal evaluation approach is temperature measurements on a hardware system. 
It can be performed using thermal imaging~\cite{sadiqbatcha2019hot, lucid_ir} or temperature sensors~\cite{zhang2021full, zhang2008chip}. 
However, the availability of the target hardware is a significant limitation. 
For example, large-scale 2.5D and 3D chiplet systems with tens of chiplets do not exist yet, while smaller prototypes and commercial systems provide limited insights applicable for larger systems~\cite{amd2021chipletTechnology,zaruba20204096}. 
This limitation motivated FEM-based modeling as the most accurate way to analyze the heat flow and temperature~\cite{sultan2019thermalSimulatorSurvey}.
% FEM analyzes heat transfer through conduction, convection, and radiation by breaking down complex systems into smaller mesh elements. 
Proprietary software, such as ANSYS Fluent~\cite{matsson2023introductionToFluent} and COMSOL~\cite{comsol}, are commonly used for FEM simulations. 
Since FEM suffers from computational cost, 
detailed FEM solutions are suitable only for small designs and validating analytical models~\cite{matsson2023introductionToFluent}. 
For example, the authors of~\cite{chipletThermalModel} employ FEM to simulate a two-chiplet system on an interposer. 
They employ abstracted FEM models for both $\mu$-bumps and C4 bumps to speed up the process, effectively reducing computational complexity before tackling the entire system. 

The computational overhead and impractically high execution time of FEM solvers 
motivate analytical models that enable rapid thermal evaluation in early design phases.
The most common method involves constructing thermal RC networks and solving the corresponding system of ODEs. 
Popular thermal simulators such as HotSpot~\cite{skadron_temperatureAware} leverage this method, focusing on the microarchitectural layout blocks to facilitate design space exploration and early-stage thermal-aware layout and placement. 
Similarly, 3D-ICE~\cite{3D_ICE} models liquid cooling with microchannels embedded between silicon layers. 
PACT~\cite{pact} also employs a similar methodology by utilizing SPICE tools as solvers, focusing on standard-cell-level thermal analysis for 2.5D systems. 
% \todo{\textit{However, these tools are not fast enough for large-scale, thermally-aware DSE of multi-chiplet 2.5D and 3D integrated systems.} 
%Reviewer wants an explanation of what fast enough means. This may be partially solved by adding runtime comparisons to 3D ice and PACT.} 
However, these tools lack essential features required for accurately modeling chiplet-based heterogeneous systems and are not fast enough for large-scale, thermally-aware design space exploration (DSE) of multi-chiplet 2.5D and 3D integrated systems, as summarized in Table~\ref{tab:related_work} and detailed next.

\begin{table}[t]
\centering
\caption{Qualitative analysis of existing analytical thermal models and our proposed thermal RC and DSS models. A detailed quantitative runtime comparison is completed in Section~\ref{sec:experimental_results}.}
\label{tab:related_work}
\resizebox{\textwidth}{!}{%
\begin{tabular}{@{}c|ccccc@{}}
\toprule
 & \textbf{\begin{tabular}[c]{@{}c@{}}Non-uniform \\ grid\end{tabular}} & \textbf{\begin{tabular}[c]{@{}c@{}}Anisotropic \\ materials\end{tabular}} & \textbf{\begin{tabular}[c]{@{}c@{}}Non-homogeneous \\ layers\end{tabular}} & \textbf{\begin{tabular}[c]{@{}c@{}}Heat dissipation from both \\ boundaries of the package\end{tabular}} & \textbf{\begin{tabular}[c]{@{}c@{}}Flexible with \\ Architecture changes\end{tabular}} \\ \midrule
\textbf{Hotspot~\cite{hotspot}} & $\times$ & $\times$ & \checkmark & \checkmark & \checkmark \\
\textbf{PACT~\cite{pact}} & $\times$ & $\times$ & \checkmark & $\times$ & \checkmark  \\
\textbf{3D-ICE~\cite{3D_ICE}} & \checkmark & $\times$ & $\times$ & $\times$\tablefootnote{With Non-uniform grid, 3D-ICE does not support secondary heat flow path.}  & \checkmark \\
\textbf{Thermal RC} & \checkmark & \checkmark & \checkmark & \checkmark & \checkmark  \\
\textbf{DSS} & \checkmark & \checkmark & \checkmark & \checkmark & $\times$ \\ \bottomrule
\end{tabular}
}
\end{table}

2.5D and 3D packages often involve materials with varying thermal conductivity across different directions.  
For example, the thermal conductivity of the C4 layer is higher in the vertical direction than in the lateral direction. 
The existing thermal models~\cite{skadron_temperatureAware,3D_ICE, pact} do not account for these anisotropic properties. 
Moreover, ~\cite{skadron_temperatureAware, pact} assume a uniform grid size for all material layers (e.g., interposer, C4 bumps, chiplets). Non-uniform grids are useful for simulating regions with substantial spatial variation in temperature without unnecessarily increasing computational costs for simulating regions of moderate temperature variation.
While 3D-ICE supports non-uniform grids for 3D architectures, it does not allow non-homogeneous layers, where different regions within the same layer can have distinct material properties. 
This limitation is particularly significant for emerging technologies, such as active interposer-based systems where chiplets are embedded~\cite{vanna2023glass}.
\textit{In contrast, our multi-fidelity models address these limitations by enabling variable thermal conductivity across different directions and allowing flexible grid sizes for each layer and block. }
Additionally, our models support heat dissipation from both package boundaries, a crucial aspect overlooked by some existing approaches. 
This capability is essential for accurately modeling advanced cooling techniques, such as immersion cooling systems and flip-chip packages, which rely on efficient heat dissipation through both package boundaries to maintain optimal performance and reliability~\cite{li2010ti,huang2023experimental}.
Table~\ref{tab:related_work} provides a detailed qualitative comparison of existing thermal modeling tools. 
% highlighting the key advantages of our analytical models over existing thermal modeling techniques.
\textit{Detailed quantitative runtime comparisons between MFIT and state-of-the-art approaches are presented in Section \ref{sec:experimental_results}.}
%As a result, a denser grid for active silicon layers and a sparser grid for others reduces the execution time without losing accuracy. 
% Finally, our thermal RC model employs an adaptive ODE solver named LSODA \cite{hindmarsh2005lsoda}, resulting in significantly faster performance compared to prior approaches (ex., RK4 ODE solver in hotspot~\cite{hotspot}).

% \todo{Worth adding short section here on thermal DSE and how MFIT could be used instead of existing thermal tools?}
Architecture-level thermal RC models are commonly used for offline studies. For example, temperature sensor placement requires using a thermal model~\cite{zanini2010temperature,chundi2017hotspot}. Similarly, thermally-aware chiplet placement techniques in \cite{tap_2.5d, thermalChipletOrganizationDarkSilicon, GIA} rely on the use of architectural thermal simulators to optimize the placement of chiplets and identify possible hotspots. While existing simulators can provide the functionality required for these studies, their prohibitive runtime prevents more extensive evaluations from being performed. Faster thermal simulation tools that do not sacrifice accuracy are required to enable more comprehensive work.

However, existing architectural simulators are inadequate for dynamic thermal and power management (DTPM). DTPM requires a much faster temperature estimation time, on the order of milliseconds, for real-time temperature management~\cite{sharifi2010accurate, Bhat2018}. 
DSS models address this need by deriving a discrete-time linear time-invariant system that models the thermal dynamics at fixed locations as a function of the power consumption. 
For instance, TILTS~\cite{TILTS} discretizes the power inputs to the chip over fixed time intervals to accelerate thermal simulations.
% A DSS model is derived for sampling frequency and system configuration. 
Hence, it needs to be reproduced when the timing requirements or the underlying hardware configuration change. The speedup gain offsets the loss of the explicit connection to the hardware parameters (e.g., thermal conductance and capacitance) and generality. 
% Our DSS model is obtained by discretizing the thermal RC model using a zero-order hold (ZOH) method. 
% It is three orders of magnitude faster than the thermal RC model it is derived from, as detailed in Section~\ref{sec:RC_to_DSS}. 
% Figure~\ref{fig:model_accuracy_vs_latency} illustrates various thermal modeling techniques with their respective execution times and accuracies.

The results of FEM, thermal RC, or other thermal simulations can also be used to train physics-informed machine learning techniques to model heat transfer in integrated circuits to reduce the thermal modeling effort~\cite{cai2021physics}. 
For example, a recent technique collects data from numeric simulations and trains a random forest model to predict the convection heat transfer coefficients for a nonlinear heat transfer problem~\cite{kwon2020machine}. 
Similarly, Hwang et al.,~\cite{hwang2018accurate} present closed-form models derived from numerical simulations for tapered micro-channels to analyze the heat transfer performance as a function of the channel geometry.
In contrast to individual classes of thermal models, this work proposes a framework to produce \textit{a family of multi-fidelity thermal models for 2.5D and 3D chiplet-based systems}. The specific set of models designed with this framework covers a wide range of accuracy and execution time trade-offs, making them suitable for different design phases. Additionally, they can be augmented by additional models, such as physics-informed ML models, with complementary accuracy and execution time trade-offs.

%\vspace{-3mm}
\section{Finite Element Method for Thermal Analysis}
\label{sec:fem_overview}
% This Section overviews how FEM models' function and the process to simulate a chiplet-based system. We also explain why FEM simulations, while extremely accurate, are inadequate for design space exploration and runtime thermal management. 

%\bh{How FEM works:}
%\subsection{FEM Operation Fundamentals}
FEM analysis begins by dividing the problem domain into small finite elements, converting the continuous governing partial differential equations (PDEs) into algebraic equations.
% The process of discretizing the problem is called meshing.
Next, the system's geometry is broken down into a lattice of small discrete cells called a ``mesh'' which approximates a larger, continuous block~\cite{matsson2023introductionToFluent}.
% The term ``mesh'' refers to a lattice of small, discrete elements used to approximate a larger, continuous block \cite{matsson2023introductionToFluent}.
After applying the PDEs and boundary conditions to each element, 
the equations are assembled into a global algebraic system, maintaining continuity between adjacent elements. This global system, representing the discretized PDEs over the whole domain, is solved numerically for the field variables at each mesh node. 
In this work,
%fluid is not modeled as explained in section \ref{ssec:fine_to_abstract_FEM}. Therefore, only 
only the equation governing solid conduction is solved~\cite{sultan2019thermalSimulatorSurvey}:
%\vspace{-1mm}
\begin{equation}
    \nabla \cdot \left[ k \nabla T \right] + \dot{q} = \rho C_v \frac{\partial T}{\partial t}
    \label{eq:heat_conduction}
\end{equation}

\noindent
where $k$ is the thermal conductivity, $T$ is the temperature, $\dot{q}$ is the heat generation rate, $\rho$ is the density, and $C_v$ is the volumetric specific heat. 
%While this simplifying assumption decreases simulation time dramatically, FEM simulation is still orders of magnitude slower our thermal modeling approach as shown in Section \ref{sec:experimental_results}.

% Overview of the steps involved in performing an FEM simulation
%\bh{The FEM pipeline:}
%\vspace{-1mm}
\subsection{Stages of the FEM Simulation Pipeline}
\begin{figure*}[ht]
    \centering
    \includegraphics[width=0.98\linewidth]{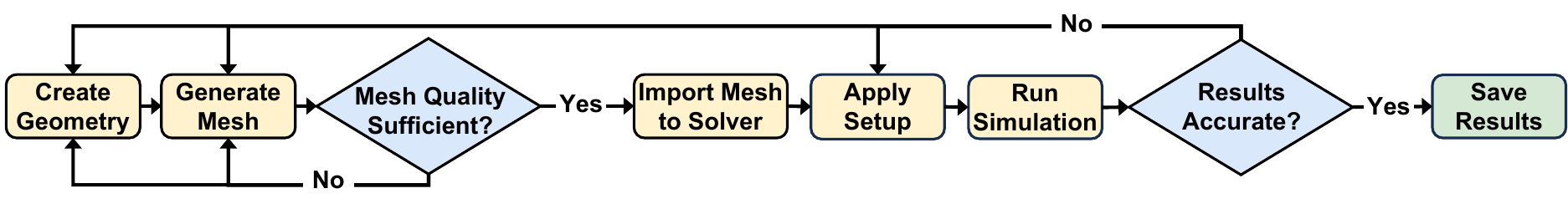}
    % \vskip -15pt
    \caption{An illustration of the FEM pipeline.}
    \label{fig:ansys_pipeline_flowchart}
    % \vskip -12pt
\end{figure*}

Performing FEM simulations involves several key processing steps visualized in \rev{Fig.}~\ref{fig:ansys_pipeline_flowchart}. 
%We highlight the three most critical stages here. 
First, the geometry, a 3D representation of the 2.5D or 3D integrated package, is created using computer-aided design tools. While increasing the detail in the geometry allows for more accuracy in modeling the system, it also increases the simulation time. This geometry should be as detailed as possible while allowing the setup and simulation to be completed within the given time constraints.
Next, a volumetric mesh is generated by transforming the 3D model into one consisting of many individual cells on which the FEM software operates.
A mesh sensitivity study is performed to determine if the mesh quality is sufficient. In this process, simulations are run with a progressively finer mesh while output parameters (such as temperature) are monitored. The mesh quality is considered to be sufficient when increasing the mesh granularity no longer impacts the temperature output.

Once an acceptable mesh has been created, it is imported into the solver. The simulation is then set up, including boundary conditions, material parameters, power source terms, and other general model parameters. 
Our specific system setup is expanded upon in Section~\ref{sec:experimental_results}. 
Finally, the FEM software simulates the model by solving the governing equations. 

%\bh{Why FEM simulation is impractical:}
%\vspace{-1mm}
\subsection{Impracticality of FEM Simulations}
While FEM simulations offer high accuracy, they are impractical for DSE or runtime thermal management due to their time-consuming setup and operation. The process of geometry creation, meshing, solver setup, and execution is intricate and often exceeds the simulation runtime itself. Because the simulation process requires multiple iterations for reliable results, this time overhead quickly becomes prohibitive. The setup of 2.5D or 3D integrated systems is especially complex due to the large number of discrete power sources. These systems also involve numerous small and large bodies, dramatically increasing the computational complexity and the solver runtime~\cite{matsson2023introductionToFluent}.
% Moreover, commercial FEM packages are {\em black-box} solvers, often necessitating heat transfer expertise to interpret results. 
Simulation times range from hours to days, directly impacted by geometric detail, complexity, size, and setup parameters such as the time step. Consequently, analyzing 2.5D or 3D integrated systems with intricate geometries and operating conditions using FEM simulations becomes prohibitively time-consuming, highlighting the need for alternative approaches.

\section{Multi-Fidelity Thermal Modeling}\label{sec:modeling_approach}
\subsection{Overview of the Proposed Approach}
\begin{figure}[b]
    \centering
    % \vskip -5pt
    \includegraphics[width=0.7\linewidth]{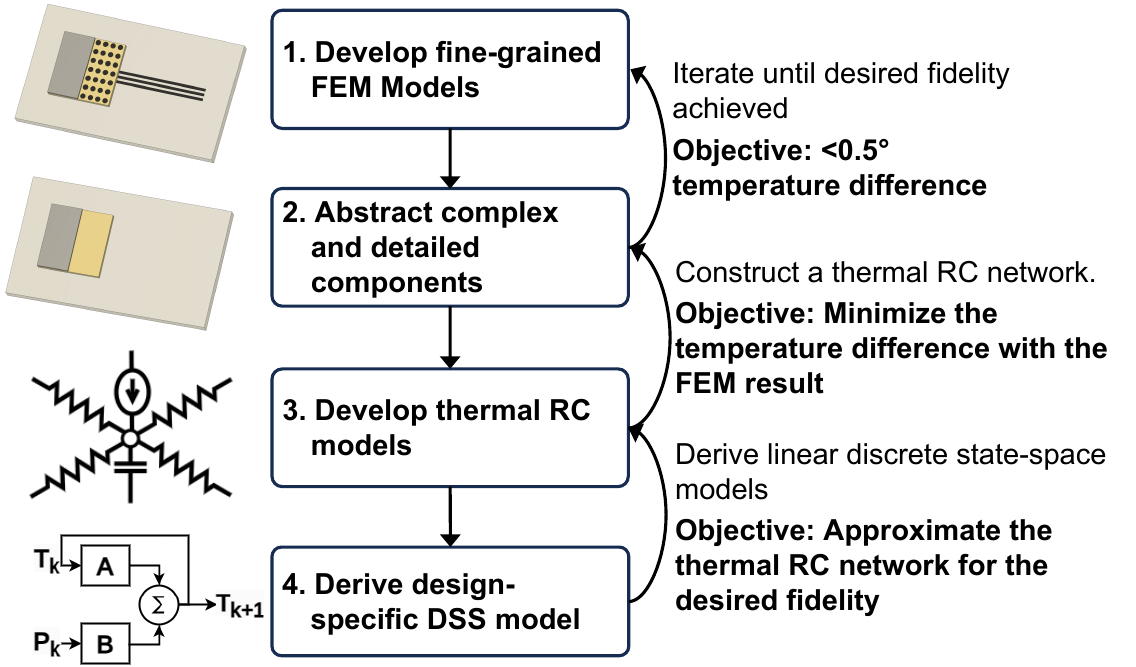}
    % \Description{...}
    % \vskip -10pt
    \caption{\rev{The proposed workflow process to produce the multi-fidelity set of thermal models, starting with the most accurate yet slow models and deriving faster models.}}
    % \vskip -15pt
    \label{fig:models_flow_diagram}
\end{figure}

Our multi-fidelity thermal model set involves four individual models, visualized in \rev{Fig.}~\ref{fig:models_flow_diagram}. The process of creating these models is identical for any packaging technology. 
% \todo{Explain new 3D geometry and cite related paper here.}
In this work, we consider 2.5D chiplet on silicon interposers and 3D systems integrated with $\mu$-bumps \cite{agarwal20223d, lau2023chiplet}. MFIT supports other integration methods and packaging technologies as well, as described below.

We start by creating fine-grained FEM models of specific components within the package. 
% These fine-grained models capture all the components in the chiplet architecture in detail. 
For example, we model individual links within the interposer and $\mu$-bumps which are used to connect chiplets to the interposer and between 3D stacked chiplets.
% connecting a chiplet to the interposer. 
The cost of this level of detail is the model complexity and execution time that limits the simulation scope.
Therefore, these fine-grained models are used as a reference to design abstracted FEM models, as explained in the following subsection. 
This abstraction enables system-level FEM simulations of systems with significantly higher chiplet counts than would otherwise be possible, with negligible accuracy loss.
All FEM simulations are performed using ANSYS Fluent~\cite{matsson2023introductionToFluent}.
The third class of model in the MFIT framework is the thermal RC model. Since these models are constructed using the geometry and material parameters of the system, a new RC model of a different system configuration can be created without re-running FEM simulations, allowing for rapid DSE.
Finally, the continuous-time state-space equations that govern the thermal RC models are discretized with a given sampling period to create DSS models as detailed in Section~\ref{sec:RC_to_DSS}.

\subsection{Fine-Grained to Abstracted FEM Modeling} \label{ssec:fine_to_abstract_FEM}

% \todo{Reviewer 1 complains that novelty of abstracted FEM is unclear. Perhaps we need to say, again, that it *is not possible* to model the system without abstracted FEM. The approach is not meant to be novel. I do not think further changes should be made to address this comment.}

First, fine-grained models of key system components are constructed with as much detail as possible. Fine-grained modeling of the entire system at the highest level of detail is infeasible due to the memory, CPU, and execution time requirements. 
The second step is systematically designing abstracted models by replacing detailed structures with homogeneous blocks. During this process, we find the material parameters for these blocks such that their thermal behavior matches the original structure.

\par
\begin{figure*}[t]
\centering
    \centering
    \includegraphics[width=\linewidth]{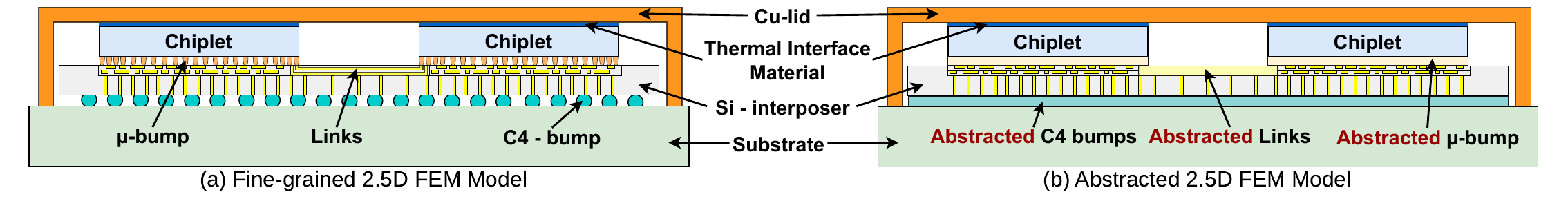}
    % \Description{...}
    % \vskip -10pt
    \caption{Cross section of a 2.5D integrated system on Si-interposer, showing abstracted blocks for the  $\mu$-bumps, C4 bumps, and link structures.}
%    \caption{Cross section of example 2 chiplet system on silicon interposer showing abstracted blocks for the  $\mu$-bumps and link structures}
    \label{fig:2.5D_cartoon_detailed_vs_abstracted}
    % \vskip -10pt
\end{figure*}

\begin{figure*}[t]
\centering
    \centering
    \includegraphics[width=\linewidth]{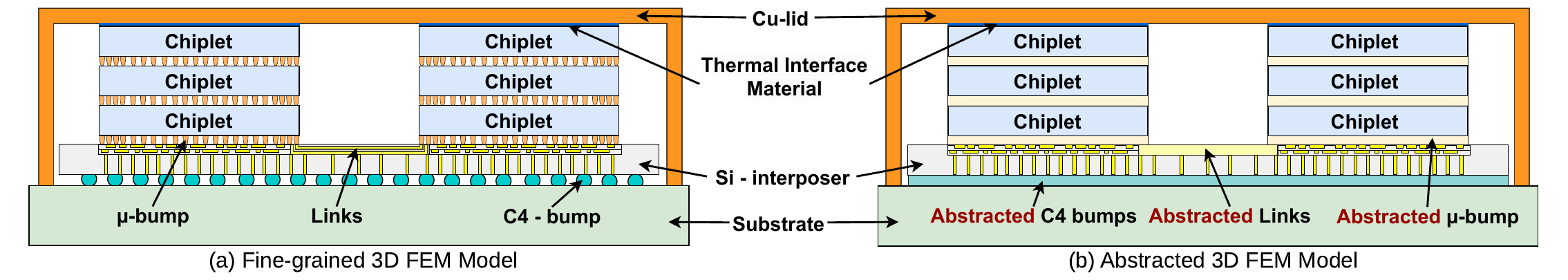}
    % \Description{...}
    % \vskip -10pt
    \caption{Cross section of a 3D integrated system, showing abstracted blocks for the $\mu$-bumps, C4 bumps and link structures.} 
    % \todo{Update explanation of figure in text to reflect new geometry.}}
%    \caption{Cross section of example 2 chiplet system on silicon interposer showing abstracted blocks for the  $\mu$-bumps and link structures}
    \label{fig:3D_cartoon_detailed_vs_abstracted}
    % \vskip -10pt
\end{figure*}

 MFIT focuses on two structures within a chiplet-based package for this work: the $\mu$-bumps connecting each chiplet to the interposer and 3D stacked chiplets to each other and the links that enable communication between chiplets. The rationale behind selecting these components is elaborated on in the following subsections. These two structures are present in both 2.5D and 3D chiplet-based packages, as shown in \rev{Figs.} \ref{fig:2.5D_cartoon_detailed_vs_abstracted} and \ref{fig:3D_cartoon_detailed_vs_abstracted}, and the results of the abstraction experiments are applied to both the 2.5D and 3D full-system abstracted models.

\par While we apply our abstract modeling approach to only two structures in this work, the same approach applies to other structures in the package, such as the substrate or C4 bumps. In addition to these abstractions, MFIT also models the heatsink as a heat transfer coefficient (HTC) instead of a physical model. This choice removes the need to model fluids in our simulations, as fluid flow in our simulations is used only for convective heat transfer in the heatsink. 

\rev{Next, we present our abstraction techniques and their validation instead of leaving them to the experimental evaluation to improve the readability and to justify larger-scale experiments performed using the abstract models.}

% Paragraph describing how the proposed approch could be used to simulate a system with a different bonding method than direct bonding.
% \todo{To capture the 3D chiplet-based systems, we consider direct die-to-die bonding between vertically stacked chiplets as an example. With this bonding method, no additional layers are modeled between stacked chiplets. Instead, chiplets are modeled directly contacting each other, stacked one on top of the other. Similarly, another bonding method can be utilized, such as TSVs or $\mu$-bumps between stacked chiplets. }
% When modeling the geometry of TSVs or $\mu$-bumps between stacked chiplets in a system, an additional block of homogeneous material is created between each stacked chiplet. This block has the same thickness as the thickness of the bonding method. Then, to determine the material parameters of this new block, an identical process is followed to that described in the following section.% , but instead utilizing the geometry of the desired bonding method.}
% \rev{To capture 3D chiplet-based systems, we consider the use of $\mu$-bumps between stacked chiplets and TSVs within chiplets \todo{\cite{}}. In this configuration, a layer of $\mu$-bumps placed between each stacked chip allows for communication.}

\subsubsection{$\mu$-bump Abstracted Model} \label{sssec:ubump_abstraction}
\begin{figure}[h!]
    \centering
    \includegraphics[width=0.7\columnwidth]{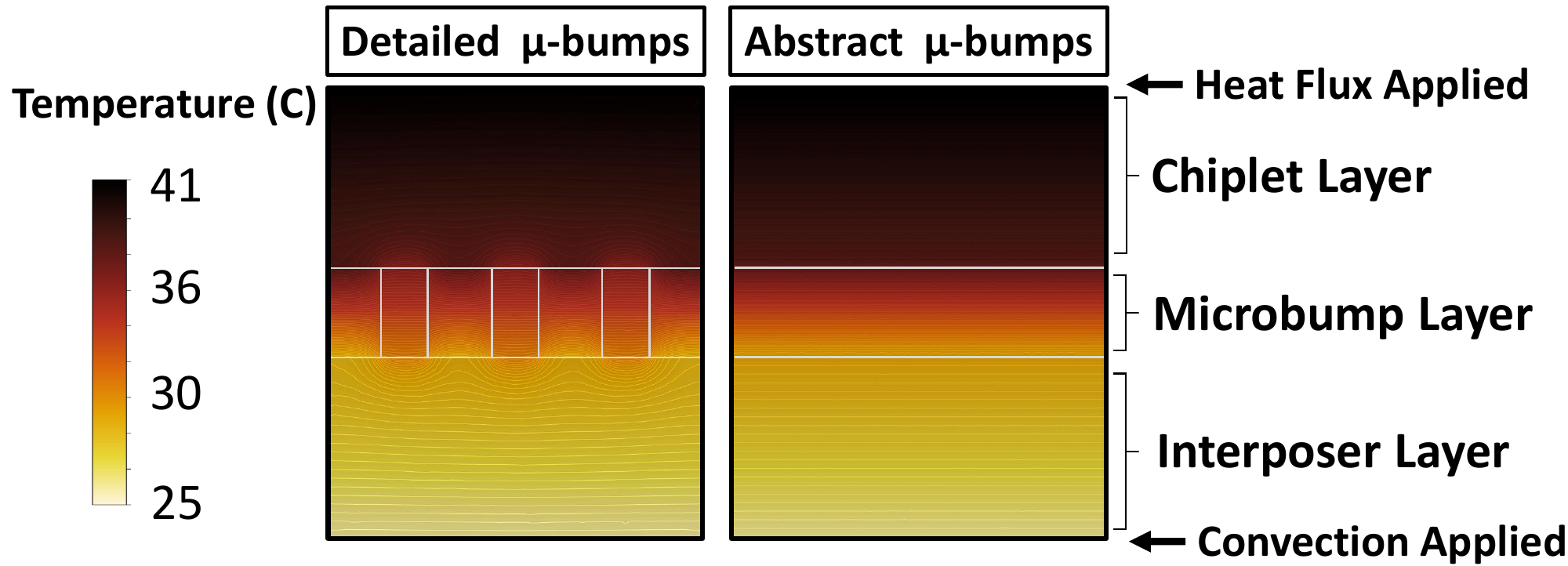}
    % \Description{...}
    % \vskip -10pt
    \caption{Temperature contour of a $\mu$-bump layer subsection.}
    \label{fig:ubump_block_detailed_contour}
    % \vskip -6pt
\end{figure}

% The $\mu$-bumps are particularly important for thermal behavior since they are one of the two paths to dissipate heat away from a chiplet, as seen in Figure~\ref{fig:2.5D_cartoon_detailed_vs_abstracted}.
% \rev{The $\mu$-bumps are particularly important for thermal behavior do to their placement in the 2.5D and 3D integrated systems. 
The $\mu$-bumps are particularly important for thermal behavior due to their placement in the 2.5D and 3D integrated systems.
In the 2.5D system, the $\mu$-bumps are one of two paths to dissipate heat from a chiplet, as illustrated in \rev{Fig.}~\ref{fig:2.5D_cartoon_detailed_vs_abstracted}. Similarly, in the 3D system, all heat flow for the lower two chiplets in each stack must travel through one of two $\mu$-bump layers, as shown in \rev{Fig.}~\ref{fig:3D_cartoon_detailed_vs_abstracted}.
Due to the density and the total number of $\mu$-bumps present, which number in the thousands, it is impractical to simulate an entire package with individually modeled $\mu$-bumps in FEM. Therefore, a small block of the $\mu$-bump layer, along with the associated upper and lower layer, is simulated in isolation to determine thermal properties that can be applied to the final abstracted models 
as illustrated in \rev{Fig.}~\ref{fig:ubump_block_detailed_contour}.

First, the detailed block containing $\mu$-bumps and underfill material is simulated with static heat flux and convection boundaries, which are applied to create a measurable thermal gradient across the $\mu$-bump layer. 
Then, the thermal conductivity $k$ is calculated as:
\begin{equation}
    k = \frac{\dot{q} \cdot l}{A \cdot \Delta T}
    \label{eq:ubump_equiv_conductivity}
\end{equation}
where $\dot{q}$ is the heat flow rate, $l$ is the thickness of the material, $A$ is the cross-sectional area, and $\Delta T$ is the temperature difference across the material
\cite{heatTransferBook}. 
Thermal capacitance and specific heat are calculated via weighted body average~\cite{bergman2011heatTransfer}. 
These parameters are applied to a model containing a homogeneous block in place of the previously modeled $\mu$-bumps and underfill material. Finally, the same boundary conditions are used, as shown in \rev{Fig.}~\ref{fig:ubump_block_detailed_contour}.

We observe identical temperature drop across the $\mu$-bump layer of the abstracted model and less than a tenth of a degree difference in interface temperatures in this sub-block, as presented in Table~\ref{table:ubump_block_results}, 
while achieving approximately 1.5x speedup. 
%
% $\mu$-bump block experiment results
\begin{table}[ht]
    %%\vspace{-6pt}
    \caption{Temperature results of the $\mu$-bump block abstraction experiments. Only a single result is shown for brevity.}
    \renewcommand{\arraystretch}{0.8}
    %%\vspace{-11pt}
    \centering
    % \resizebox{\columnwidth}{!}{
        \begin{tabular}{@{}c|ccc@{}}
            \toprule
            \textbf{Model} & \textbf{\begin{tabular}[c]{@{}c@{}}Upper surface \\ Temp. ($^{\circ}$C)\end{tabular}} & \textbf{\begin{tabular}[c]{@{}c@{}}Lower surface \\ Temp. ($^{\circ}$C)\end{tabular}} & \textbf{\begin{tabular}[c]{@{}c@{}}Temp. \\ Drop ($^{\circ}$C)\end{tabular}} \\ 
            \midrule
            \textbf{Detailed $\mu$-bumps} & 39.13 & 31.05 & 8.08 \\
            \textbf{Abstracted $\mu$-bumps} & 39.26 & 31.18 & 8.08 \\ 
            \bottomrule
        \end{tabular}
    % }
    %%\vspace{-9pt}
    \label{table:ubump_block_results}
\end{table}

% % Temperature contour comparison of the link vs no link experiment
% \begin{figure}[t]
%     \centering
%     \includegraphics[width=0.45\textwidth]{figures/ubump_experiments/ubump_block_detailed_contour.png}
%     \caption{Temperature Contour of a subsection of the $\mu$-bump layer.}
%     \label{fig:ubump_block_detailed_contour}
% \end{figure}

\subsubsection{Link Abstracted Model}
% A link is a group of wires embedded in the interposer to interconnect chiplets. Depending on the thermal crosstalk over the links, the NoI architecture can significantly affect a system's thermal behavior. To determine how links are modeled in the complete system simulations, we tested three different configurations of a two-chiplet package. 
% These configurations model the link (1) in full detail, (2) as an abstracted block, and (3) not included at all. 
\rev{A link refers to a group of wires embedded in the interposer that interconnect chiplets. Depending on the thermal crosstalk through these links, the Network-on-Interposer (NoI) architecture can influence the overall thermal behavior of a system. To understand the significance of accurately modeling links in full-system thermal simulations, we evaluated three configurations for a two-chiplet package:}

\rev{\noindent \textbf{1. Full-detail link model:} This configuration explicitly models each horizontal and vertical wire segment in the interposer layer, including their placement and density between the chiplets. Specifically, we implemented two $\times$64 Advanced Package Lane interfaces, following the bump map layout described in the UCIe specification~\cite{sharma2022universal}. This detailed model captures the realistic routing and thermal characteristics of Redistribution Layer (RDL)-based interposer connections.}

\rev{\noindent \textbf{2. Abstracted block model:} Here, the link is represented as a single thermal block with averaged properties, simplifying its spatial granularity.}

\rev{\noindent \textbf{3. No-link model:} In this case, the link is completely omitted from the thermal model.}

We use two different power configurations: (a) the power dissipation is static over time, and (b) it varies dynamically over time. 
These power consumption profiles are applied to one chiplet while the temperature of the other chiplet is calculated through FEM simulation. 
\rev{The mean absolute error (MAE) and average percentage error of the receiving chiplet temperature compared to the detailed model case are recorded in Table~\ref{table:link_experiments_accuracy}. 
Only a minimal accuracy loss is observed for both cases}, 
while execution time savings are significant, as shown in Table~\ref{table:link_experiment_execution_time}. 
Therefore, we choose not to model links in our full-system simulations. 

% Table with a temperature difference comparison of abstracted and no links vs. the case of the detailed link
\begin{table}[ht]
    % %\vspace{-8pt}
    \caption{\rev{MAE and Average Percentage Error comparison between abstracting and removing the links compared to detailed link modeling while operating between 60$^{\circ}$C and 100$^{\circ}$C.}}
    %\vspace{-8pt}
    \renewcommand{\arraystretch}{0.8}
    \centering
    \begin{tabular}{@{}c|cccc@{}}
    \toprule
    \textbf{\begin{tabular}[c]{@{}c@{}} Power \\ ~ \end{tabular}} & \textbf{\begin{tabular}[c]{@{}c@{}}Steady-State \\ MAE ($^{\circ}$C)\end{tabular}} & \textbf{\begin{tabular}[c]{@{}c@{}}\rev{Steady-State} \\ \rev{Avg. \% Error}\end{tabular}}& \textbf{\begin{tabular}[c]{@{}c@{}}Transient \\ MAE ($^{\circ}$C)\end{tabular}} & \textbf{\begin{tabular}[c]{@{}c@{}}\rev{Transient} \\ \rev{Avg. \% Error}\end{tabular}} \\ \midrule
    \textbf{Abstracted links} & 0.05$^{\circ}$C & \rev{0.083\%} & 0.02$^{\circ}$C & \rev{0.045\%} \\
    \textbf{No links} & 0.34$^{\circ}$C & \rev{0.429\%} & 0.13$^{\circ}$C & \rev{0.259\%} \\ \bottomrule
    \end{tabular}
    %\vspace{-10pt}
    \label{table:link_experiments_accuracy}
\end{table}

% Table showing the runtime of the Detailed, Abstract, and No Links experiments in Transient and Steady solver.
\begin{table}[ht]
\caption{Execution time for Detailed, Abstract, and No Links experiments with steady and transient power inputs.}
%\vspace{-8pt}
\centering
\renewcommand{\arraystretch}{0.8}
\begin{tabular}{@{}c|cc@{}}
\toprule
\textbf{\begin{tabular}[c]{@{}c@{}} Power \\ ~ \end{tabular}} & \textbf{\begin{tabular}[c]{@{}c@{}}Steady\\ Exe. time (min)\end{tabular}} & \textbf{\begin{tabular}[c]{@{}c@{}}Transient \\ Exe. time (min)\end{tabular}} \\ \midrule
\textbf{Detailed links} & 489.23 & 503.86 \\
\textbf{Abstract links} & 164.29 & 172.64 \\
\textbf{No links} & 123.80 & 132.13 \\ \bottomrule
\end{tabular}
%\vspace{-10pt}
\label{table:link_experiment_execution_time}
\end{table}

\rev{\subsubsection{TSV Abstracted Model}
Silicon interposers with through silicon vias (TSVs) are one of the prominent methods for integrating multiple chiplets into a single package. In silicon interposers and 3D stacked chips, TSVs are crucial structures to conduct signals and power through the chip to layers above and below. We develop and validate abstracted TSV models similar to the $\mu$-bump modeling presented in Section \ref{sssec:ubump_abstraction}.}

\rev{The first step is modeling a detailed block of the TSV layer, including the TSVs themselves and the surrounding silicon material. Additional blocks are placed above and below this layer to represent adjacent layers present in the full system. To model the thermal conductivity of the TSV later, heat flux and convection are applied to the top and bottom blocks, respectively. This standard process \cite{heatTransferBook} a temperature difference $\Delta$T across the TSV layer, as visualized in Fig. 7 for the u-bump layer. Then, Equation \ref{eq:ubump_equiv_conductivity} is used to find the effective thermal conductivity $k$.}

\rev{The second step is to create the abstracted thermal model, where the TSV layer is converted from a detailed representation to a single homogeneous block. The effective thermal conductivity found using the detailed model above is applied to this homogeneous block. Then, the same heat flux and convection which were applied to the detailed model are applied to this abstracted geometry. Finally, the temperature difference is measured again at the top and bottom of the TSV layer and compared to the temperature difference found using the detailed model.}

\rev{The results of these experiments are shown in Table \ref{tab:tsv} for two separate TSV configurations. The first configuration considers 15$\mu$m diameter TSVs with a pitch of 25$\mu$m \cite{yu2011fabrication}. The second one considers 5$\mu$m diameter TSVs with a pitch of 10$\mu$m \cite{satheesh2012power}. Results show that the temperature drop across the TSV layer is within 0.5 $^{\circ}$C for both the large and small TSV diameter cases. Additionally, the speedup from modeling the detailed to abstract TSV case is approximately 1.8x, in line with the $\mu$-bump abstraction speedup.}

\begin{table}[h]
\centering
\renewcommand{\arraystretch}{0.8}
\caption{\rev{Temperature results of the TSV block abstraction experiments. Results are shown} }
\label{tab:tsv}
\begin{tabular}{@{}c|c|ccc@{}}
\toprule
\textbf{\rev{TSV type}} & \textbf{\rev{Model}} & \textbf{\begin{tabular}[c]{@{}c@{}}\rev{Top surface} \\ \rev{temperature ($^{\circ}$C)}\end{tabular}} & \textbf{\begin{tabular}[c]{@{}c@{}}\rev{Bottom surface} \\ \rev{Temperature ($^{\circ}$C)}\end{tabular}} & \textbf{\begin{tabular}[c]{@{}c@{}}\rev{Temperature} \\ \rev{Drop ($^{\circ}$C)}\end{tabular}} \\ \midrule
\multirow{2}{*}{\textbf{\rev{15$\mu$m diameter, 25$\mu$m pitch}}} & \textbf{\rev{Detailed}} & \rev{85.03} & \rev{60.93} & \rev{24.10} \\
 & \textbf{\rev{Abstracted}} & \rev{84.77} & \rev{61.18} & \rev{23.59} \\ \midrule
\multirow{2}{*}{\textbf{\rev{5$\mu$m diameter, 10$\mu$m pitch}}} & \textbf{\rev{Detailed}} & \rev{85.95} & \rev{61.11} & \rev{24.84} \\
 & \textbf{\rev{Abstracted}} & \rev{85.87} & \rev{61.18} & \rev{24.69} \\ \bottomrule
\end{tabular}%
\end{table}

\subsubsection{Heatsink Abstracted Model}
\label{ssec:heatsink}
FEM simulations that involve a heatsink must model the convective heat transfer from the system to the atmosphere using fluid models~\cite{chipletThermalModel}. However, modeling the fluid dynamics dramatically increases the setup and runtime of FEM simulations. 
Additionally, the geometry must be modified for every different heatsink configuration, further increasing the time needed for design iteration. Due to their complexity, high-performance cooling methods, such as liquid cooling, are also difficult to model in FEM-based simulations.

In MFIT, we remove the need to model the heatsink by abstracting the cooling solution to a single HTC. This coefficient is applied to the top of the lid where a heatsink is typically attached. Modeling a heatsink as a HTC is a common practice for many different cooling solutions \cite{khan2008cylindricalPinHeatsink, narasimhan2002heatsinkModels}. This approach allows for a great deal of flexibility in FEM modeling. Instead of completing the time-consuming pipeline in \rev{Fig.}~\ref{fig:ansys_pipeline_flowchart}, the HTC can be easily modified to test the behavior of different cooling solutions.

The experiments demonstrated in this work assume an active air-cooled heatsink. The value of the HTC of an air-cooled heatsink is determined by:
\begin{equation}
    h_{eq} = \frac{h_{avg} \cdot A_t \cdot \left( 1 - \frac{N \cdot A_f \cdot (1 - \eta_f)}{A_t} \right)}{LW}
    \label{eq:heatsink_convection_coef}
\end{equation}
where $A_t$ is the total area of the heatsink, $A_f$ is the fin area, $N$ is the number of fins, $\eta_f$ is the fin efficiency, and $L$ and $W$ are the length and width of the base plate. The average convective HTC ($h_{avg}$) can be calculated using the Nusselt number~\cite{bergman2011heatTransfer}. We select values consistent with a basic copper heatsink with forced airflow provided by a typical commercial computer fan. 

% \todo{Explain that traditional liquid cooling can be modeled easily with HTC. Add equation for heat transfer in liquid cooling system.}

While an active air-cooled heatsink is assumed in this work, similar equations to Equation~\ref{eq:heatsink_convection_coef} for liquid cooling exist~\cite{bergman2011heatTransfer}. If the system designer chose to use liquid cooling for the simulation, a new HTC would be calculated and applied in place of the air cooling HTC described previously.
% The parameters can be easily tuned to reflect different cooling solutions. }
%, providing an approximately 3x speedup.

While the above method of abstracting a heatsink is effective for systems where heat is dissipated primarily through the external boundaries of the package (top of the lid, bottom of the substrate), a different modeling approach may be used for other cooling methods, such as inter-tier liquid cooling, where microchannels contact the chip directly~\cite{hwang2018accurate, kwon2020machine}. In such an approach, heat is dissipated directly from the chip without moving to a heat spreader like the lid. Hence, additional heat transfer coefficients or abstraction techniques may be needed to capture the cooling behavior. 

% FEM to thermal RC
\subsection{FEM to Thermal RC models} \label{ssec:FEM_to_RC}
% Explain the process

% \todo{Reviewer 1 thinks it is unclear how the FEM model is transformed into a RC model, such as what tools are used. Also need better explanation for how and why capacitance is tuned.}

This section describes the process of constructing a thermal RC model from the geometry of a given package. MFIT applies this technique to both 2.5D and 3D systems demonstrating the flexibility of the proposed methodology to different packaging technologies. However, this process can easily be applied to any package. Table \ref{tab:notations} provides the notation of the equations used in this and the following subsection.

\begin{table}[b]
  \centering
  \caption{\rev{Summary of the key parameters and notations used in this work.}}
  \label{tab:notations}
  \begin{tabular}{@{}c|l@{}}
    \toprule
    \textbf{Notation}                  & \textbf{Definition} \\ \midrule
    $k_x$, $k_y$, $k_z$                & Thermal conductivity of a layer along the x, y, and z axes \\
    $G_x$, $G_y$, $G_z$                & Thermal conductance of a node along the x, y, and z axes \\
    $\rho$                           & Material density \\
    $C_v$                            & Volumetric specific heat of a layer \\
    $G_{conv}$                       & Convection conductance of a boundary node \\
    $h_{eq}$                         & Heat transfer coefficient of heatsink \\
    N                                & Total number of nodes in the RC and DSS models \\
    $\mathbf{T}$, $\dot{\mathbf{T}}$  & Nx1 temperature matrix and its derivative \\
    $\dot{\mathbf{q}}$               & Nx1 matrix of heat generation \\
    $\mathbf{C}$, $\mathbf{G}$       & NxN thermal capacitance and conductance matrices \\
    \bottomrule
  \end{tabular}
\end{table}

% \todo{\textbf{Description here of how inputs (geometry, materials, etc.) are supplied to the RC solver}}

% This section describes the construction of the thermal RC models using the chiplet-based package geometry and material properties.
The package is first divided into horizontal layers, with the slicing process starting at the bottom substrate layer and ending at the top lid layer. 
Depending on the package design, each layer may be composed of a uniform material or various material blocks, resulting in either homogeneous or \textit{non-homogeneous layers}.
% Non-homogeneous layers consist of different materials with distinct thermal properties.}
% as detailed in \todo{Table~\ref{tab:related_work}}.}
This flexibility enables the thermal RC model to simulate packages with heterogeneous designs where different chiplets are manufactured with various technologies, resulting in different material parameters in the same layer.
% allows for different computing elements within the same layer to be made using various technologies, resulting in diverse material blocks.
%as shown in Figure~\ref{fig:cartoon_detailed_vs_abstracted}(b).
Layers with uniform material properties are divided into a 2D grid of nodes, where the grid granularity can vary between layers. For non-homogeneous layers with different material blocks, each block can have a distinct grid granularity, resulting in a \textit{non-uniform grid} that connects the entire package as a 3D network of thermal nodes, discretizing the chiplet geometry in space.

Since a layer or material blocks may have \textit{anisotropic material}, where thermal conductivity differs along the x, y, and z axes (represented by $k_x$, $k_y$, and $k_z$), we calculate the thermal conductance ($G_x$, $G_y$, $G_z$) for each node using the following equations:
\begin{equation}
    G_x = \frac{k_x\cdot l_y\cdot l_z}{l_x} ,~
    G_y = \frac{k_y\cdot l_x\cdot l_z}{l_y} ,~
    G_z = \frac{k_z\cdot l_x\cdot l_y}{l_z}
    \label{eq:conductance}
    %\vspace{-1mm}
\end{equation}
where $l_x$ and $l_y$ are the node lengths in the x and y dimensions, respectively, and $l_z$ represents the thickness of the layer. 
The thermal capacitance of each node is then calculated as $C = \rho\cdot C_v \cdot l_x \cdot l_y \cdot l_z$, where $\rho$ is material density and $C_v$ is the volumetric specific heat.

%, given by the following equation:
% %
% \begin{equation}
%     C = \rho\cdot C_v \cdot l_x \cdot l_y \cdot l_z
%     \label{eq:capacitance}
% \end{equation}
% %

Heat is dissipated from the package primarily through a heatsink which is simulated using a convective heat transfer coefficient as detailed in Section~\ref{ssec:heatsink}.
MFIT assumes forced convection is applied to the heatsink while passive convection occurs on the other external boundaries of the package. 
Consequently, convective conductance ($G_{conv} = h_{eq} \cdot l_x \cdot l_y$) is incorporated into the nodes of the top and bottom layers.
% computed using the following equation:
% %
% \begin{equation}
%     G_{conv} = h_{eq} \cdot l_x \cdot l_y
%     \label{eq:conv_conductance}
% \end{equation}
%

% Once all thermal capacitance and conductance values are found, all nodes are connected, forming a 3D RC network. This network leads to a system of ODEs that can be represented in continuous state-space form, as shown in the following equation:
The conductance between neighboring nodes $i$ and $j$ ($G_{ij}$) of the same layers is determined by lateral conductance ($G_x$ and $G_y$).
Since our thermal RC model allows non-uniform grid sizes for different layers and blocks, a node in one layer can be connected to multiple nodes from adjacent layers. 
Thus, vertical conductance between nodes of different layers is calculated from $G_z$, considering the overlap in the x-y plane.
Once this RC network is established, we can formulate an ODE based on Kirchhoff's current law for a node $i$ as:
\setlength{\abovedisplayskip}{6pt}
\setlength{\belowdisplayskip}{6pt}
\begin{equation}
   C_i \frac{dT_i}{dt} = \textstyle\sum_{j=1}^{N} (G_{ij}) (T_j - T_i) + \dot{q}_i 
\label{eq:ode}
\end{equation}
\begin{align*}
\text{where}~~G_{ij} = 
    0, ~\text{if } i == j \text{ or } j \text{ is not a neighbor node of } i.
\end{align*}

The heat generation ($\dot{q}_i$) from node $i$ is analogous to electric current, and temperature ($T_i$) is analogous to voltage. 
Since only the chiplet layers consume power, heat generation for the nodes in other layers is zero. %(\dot{q}_i$) is zero.
Solving the system of ODEs by forming a matrix is a well-studied approach. It can be represented by:
\begin{equation}
    \mathbf{C} \times \dot{\mathbf{T}} = \mathbf{G} \times \mathbf{T} +  \dot{\mathbf{q}}
    \label{eq:ode_matrix}
\end{equation}
where $\mathbf{T}$, $\dot{\mathbf{T}}$, and $\dot{\mathbf{q}}$ are $N \times 1$ matrices representing node temperatures, temperature derivative, and generated heat.
$\mathbf{C}$ is a $N\times N$ diagonal matrix, where each element corresponds to a node's thermal capacitance.
The conductance matrix $\mathbf{G}$ can be expressed as:
% is constructed by rearranging $G_{ij}$'s from Equation~\ref{eq:ode} as follows:
\begin{equation}
    \mathbf{G} = \begin{bmatrix}
    -\sum_{j=1}^{N} G_{1j} & G_{12} & \cdots & G_{1N} \\
    G_{21} & -\sum_{j=1}^{N} G_{2j} & \cdots & G_{2N} \\
    \vdots & \vdots & \ddots & \vdots \\
    G_{N1} & G_{N2} & \cdots & -\sum_{j=1}^{N} G_{Nj}
    \end{bmatrix}
    \label{eq:g_matrix}
\end{equation}
where, $G_{ij}$ represents conductance between the neighboring nodes $i$ and $j$.
% which can be derived from $G_x$, $G_y$ and $G_z$.

% \begin{equation}
%     \mathbf{G} = \begin{bmatrix}
%     -\sum_{j\neq 1} G_{1j} & G_{12} & \cdots & G_{1N} \\
%     G_{21} & -\sum_{j\neq 2} G_{2j} & \cdots & G_{2N} \\
%     \vdots & \vdots & \ddots & \vdots \\
%     G_{N1} & G_{N2} & \cdots & -\sum_{j\neq N} G_{Nj}
%     \end{bmatrix}
%     \label{eq:g_matrix}
% \end{equation}

% MFIT employs the highly adaptive solver LSODA~\cite{hindmarsh2005lsoda} %\footnote{We used C library from: \url{https://github.com/sdwfrost/liblsoda}; the original implementation was written in Fortran.}
% to solve this system of ODEs. 
% LSODA is designed to handle both stiff and non-stiff systems efficiently. 
% It dynamically switches between different numerical integration methods depending on the characteristics of the ODE system being solved. 
% This switching capability is particularly useful for thermal ODEs, as the equations' stiffness can vary over time depending on the power consumption. 
% It is worth noting that the matrices representing the system are highly sparse since each node is connected to only a few neighboring nodes. 
% MFIT leverages this sparsity to accelerate the solver's execution time. 
% \todo{Need to explain new solver and how/why capacitance is tuned.}

MFIT employs the backward Euler method to solve this system of ODEs, as its implicit formulation ensures stability when handling the stiff equations typical of thermal modeling. 
Given that the system's matrices are highly sparse (each node connects to only a few neighbors), MFIT leverages this sparsity by integrating SuperLU~\cite{li2005overview} as a sparse linear solver. 
Additionally, we incorporate the BLAS library \cite{ReferenceBLAS} alongside SuperLU to further accelerate computations. 
To balance flexibility and performance, we implement the front-end in Python to define the geometry and ODE Equations, while the core computations rely on optimized C-based implementations of SuperLU.

% Finally, we fine-tune the capacitance values of each layer utilizing FEM results as a reference to improve the accuracy of our model.
\bh{Capacitance Tuning:}
While the thermal RC method provides very accurate steady-state temperature predictions, transient temperature calculations can exhibit relatively higher errors. 
This is a limitation of existing approaches as discussed in the experimental evaluations. %, as shown in Table~\ref{tab:combined_accuracy}. 
To mitigate this issue, we fine-tune the capacitance values of each layer in the system using FEM results as a reference. 
Specifically, we introduce a scalar multiplier for each layer’s capacitance and optimize these values using the nonlinear optimizer Nelder-Mead~\cite{barton1996nelder}.
This tuning process is performed on a small-scale system and is based on the number of layers and their material properties rather than chiplet placement or geometry. 
Once optimized, the same capacitance parameters are applied to larger system sizes for more accurate thermal modeling without increasing grid complexity, as we demonstrate in Section~\ref{ssec:accuracy}. 
% For tuning, we use the 2×2 chiplet system for all 2.5D configurations and the 2×2×3 chiplet system for 3D configurations. 
Since most advanced packaging technologies rely on similar materials, \textit{re-tuning is rarely required}. 
Tuning is only necessary when the number of layers or material properties changes, in which case fine-tuning capacitance values on a small-scale system ensures accurate temperature calculation. In this work, we tune two representative small-scale systems, one for 2.5D systems and one for 3D system. The capacitance values of these representative systems are applied to the larger systems shown in the experimental results. 

%\vspace{-1mm}
\subsection{Thermal RC to Discrete State Space models}
\label{sec:RC_to_DSS}
% Explain how the RC model is discretized to create a discrete state space model
% This model is based on the same C and G matrix determined in the RC model.
% Therefore, creating the Discrete model without first creating the RC model is impossible.
% The Discrete model is as accurate as the RC model but runs much faster.
The thermal RC model can be discretized in the time domain to further reduce the execution time of the model with no cost of accuracy. The discretization of a thermal RC model is completed by automated tools and takes only milliseconds to run. 
The DSS model can be used for dynamic thermal management applications where the geometry is fixed, and faster thermal prediction is more important. 
However, a limitation of the DSS model is its dependence on an underlying continuous time model. It cannot be constructed directly without a thermal RC model as an intermediate step or system identification and measurement data. Additionally, a DSS model is specific to the geometry, materials, and sampling period used during the creation of the thermal RC model and the later discretization process. Therefore, the DSS model must be reconstructed if any design parameter changes. Only an existing RC model and a set time step are required to create a DSS model, with no direct information from the previous FEM model being needed. 

\begin{comment}
Integrating the continuous time system, as given in Equation~\ref{eq:ode_matrix} for $t=t+T_s$ results in~\cite{zak2003systems}:
\begin{equation}
\begin{split}
    \mathbf{T}(t + T_s) = & e^{\mathbf{C^{-1}\mathbf{G}}(T_s)}~\mathbf{T}(t) + \int_{t}^{t+T_s}e^{\mathbf{C^{-1}\mathbf{G}}(t+T_s - \tau)} {C^{-1}}(\tau)~\dot{\mathbf{q}}~d\tau
\end{split}
\label{eq:int}
\end{equation}
%
\todo{Add more in depth explanation of DSS model (equations) between 8 and 9} Assuming constant heat generation during the sampling period $T_s$, the system can be discretized from continuous time variable $t$ to discrete steps $k$ as:
% In this work, the thermal RC model is discretized from continuous time to discrete time using the zero-order hold (ZOH) method for a given sampling period $T_s$. 
% ZOH matches the continuous time model when power is provided at each sampling period as discrete inputs. A larger or smaller sampling period can be used for the discretization process depending on the sampling period of the required power input. The resulting DSS model takes the form:
% \begin{equation}
%     \mathbf{T}[k+1] = (I - T_s \mathbf{C}^{-1} \mathbf{G}) \mathbf{T}[k] + T_s \mathbf{C}^{-1} \dot{\mathbf{q}}[k]
% \end{equation}
% where $I$ is the identity matrix. This equation can be simplified to the following form:
\begin{equation}
    \mathbf{T}[k+1] = \mathbf{A}\times \mathbf{T}[k] + \mathbf{B}\times \dot{\mathbf{q}}[k]
    \label{eq:DSS_model}
\end{equation}
where $\mathbf{A}$ and $\mathbf{B}$ are the state and input matrices. 
\end{comment}

Rearranging the Equation~\ref{eq:ode_matrix} into a state-space representation:
\begin{equation}
    \dot{\mathbf{T}} = \mathbf{A} \mathbf{T} + \mathbf{B} \dot{\mathbf{q}}, \quad \text{where } \mathbf{A} = \mathbf{C}^{-1} \mathbf{G}, \quad \mathbf{B} = \mathbf{C}^{-1}.
\end{equation}

The time evolution of this system over an interval $t \in [kT_s, (k+1)T_s]$, where $T_s$ denotes the sampling period, is determined by solving the differential equation. 

In the absence of power ($\dot{\mathbf{q}} = 0$), the homogeneous solution is given as:
\begin{equation}
    \mathbf{T}(t) = e^{\mathbf{A} (t - kT_s)} \mathbf{T}(kT_s)
    \label{eq:dss_homogeneous}
\end{equation}

Assuming $\dot{\mathbf{q}}$ is present and remains constant over the interval under a zero-order hold (ZOH) approximation, the particular solution is given as:
\begin{equation}
    \mathbf{T}(t) = \int_{kT_s}^{t} e^{\mathbf{A}(t - \tau)} \mathbf{B} \dot{\mathbf{q}}(\tau) \, d\tau
    \label{eq:dss_perticular}
\end{equation}
Combining Equations~\ref{eq:dss_homogeneous} and ~\ref{eq:dss_perticular}:
\begin{equation}
    \mathbf{T}(t) = e^{\mathbf{A} (t - kT_s)} \mathbf{T}(kT_s) + \int_{kT_s}^{t} e^{\mathbf{A} (t - \tau)} \mathbf{B} \, d\tau \cdot \dot{\mathbf{q}}(kT_s)
\end{equation}
Evaluating at $t = (k+1)T_s$, the system is discretized as:
\begin{equation}
    \mathbf{T}((k+1)T_s) = e^{\mathbf{A} T_s} \mathbf{T}(kT_s) + \left( \int_{0}^{T_s} e^{\mathbf{A} \tau} d\tau \right) \mathbf{B} \dot{\mathbf{q}}(kT_s), \quad \text {where} \int_{0}^{T_s} e^{\mathbf{A} \tau} d\tau = \mathbf{A}^{-1} (e^{\mathbf{A} T_s} - \mathbf{I})
\end{equation}
Then, the discrete-time system matrices is defined as:
\begin{equation}
    \mathbf{A_d} = e^{\mathbf{A} T_s}, \quad \mathbf{B_d} = \mathbf{A}^{-1} (\mathbf{A_d} - \mathbf{I}) \mathbf{B}
\end{equation}
the resulting discrete-time state-space equation is given by:
\begin{equation}
    \mathbf{T}[k+1] = \mathbf{A_d} \mathbf{T}[k] + \mathbf{B_d} \dot{\mathbf{q}}[k]
    \label{eq:DSS_model}
\end{equation}

Equation~\ref{eq:DSS_model} represents the discrete-time equivalent of the continuous-time thermal RC model (shown in Equation~\ref{eq:ode_matrix}).
MFIT uses the zero-order hold (ZOH) method as given in Equation~\ref{eq:dss_perticular}  for the discretization process.
When power is provided as discrete inputs at each sampling period, ZOH provides an exact match to the continuous time model. 
$T_s$ can be determined for discretization as a function of input power consumption and system dynamics.

\textit{The DSS model consists only of multiply-accumulate operations, allowing for extremely fast operation, as shown in Section \ref{sec:experimental_results}. The discretization process has only a one-time cost and is also nearly instantaneous, allowing for rapid DSS model creation when a thermal RC model is available. }

\section{Experimental Results} \label{sec:experimental_results}
% This section will go over the accuracy and performance of the thermal RC and DSS model, a comparison to the existing simulator HotSpot, and case studies

% This section will show details of the setup
\subsection{Experimental Setup}\label{ssec:setup}
We evaluate the accuracy of the proposed MFIT methodology on three 2.5D systems and one 3D system representative of their respective classes. Three separate 2.5D systems are studied to demonstrate the flexibility of the proposed approach for systems with different numbers of chiplets. A 3D system is considered to demonstrate the capability of the approach to model systems beyond a single planar layer of chiplets. 
\textit{Our thermal RC and DSS models are open-sourced to catalyze research in this domain.}
The rest of this section describes the parameters and geometry of the 2.5D and 3D systems considered in this paper.
%Details of the parameters of the 2.5D and 3D system are listed in Table \ref{tab:system_configurations}.

\begin{table}[b!]
\caption{\rev{Specifications of simulated systems in this work.}}
\label{tab:system_configurations}
% \resizebox{\columnwidth}{!}{
\begin{tabular}{@{}ccccc@{}}
\toprule
\multicolumn{1}{c|}{\textbf{Parameter}} & \textbf{16 2.5D} & \textbf{36 2.5D} & \textbf{64 2.5D} & \textbf{16x3 3D} \\ \midrule
\multicolumn{5}{c}{\textbf{\rev{Package Geometry}}} \\ \midrule
\multicolumn{1}{c|}{Package Thickness (mm)} & 1.855 & 1.855 & 1.855 & 2.105 \\
\multicolumn{1}{c|}{Package Length and Width (mm)} & 15.5 & 21.5 & 27.5 & 15.5 \\
\multicolumn{1}{c|}{Package Top Area (mm$^2$)} & 240.25 & 462.25 & 756.25 & 240.25 \\
\multicolumn{1}{c|}{Package Volume (mm$^3$)} & 445.66 & 857.47 & 1402.84 & 505.72 \\ \midrule
\multicolumn{5}{c}{\textbf{\rev{Power (100\% utilization)}}} \\ \midrule
\multicolumn{1}{c|}{\rev{Individual Chiplet Maximum Power (W)}} & 3 & 3 & 3 & 1.2 \\
\multicolumn{1}{c|}{\rev{Total System Maximum Power (W)}} & 48 & 108 & 192 & 57.6 \\
\multicolumn{1}{c|}{\rev{Total System Maximum Power per lid area (W/mm$^2$)}} & 0.199 & 0.233 & 0.253 & 0.239 \\ \midrule
\multicolumn{5}{c}{\textbf{\rev{Temperature}}} \\ \midrule
\multicolumn{1}{c|}{Maximum Chiplet Temperature ($^{\circ}$C)} & 118.25 & 129.75 & 164.03 & 142.01 \\ \bottomrule
\end{tabular}
% }
\end{table}

\subsubsection{Package Overview}
Both the 2.5D and 3D systems utilize a silicon interposer with chiplets placed upon it. The interposer is connected to the underlying substrate using C4 bumps. A copper lid covers the chiplets and acts as a heat spreader. The lid and substrate form the outer bounds of the package. Dimensions of the package for each system configuration are found in Table \ref{tab:system_configurations}. Within the package, copper wires embedded in the interposer are used to connect neighboring chiplets. In both our 2.5D and 3D systems, each chiplet area is considered to be $2.25 mm^2$, consistent with prior studies~\cite{krishnan2021siam,sharma2022swap,sharma2023florets}. 
Each chiplet consists of multiple blocks. Each of these blocks corresponds to a component, such as a computational tile or a router used for inter-chiplet communication, as detailed in \rev{Fig.}~\ref{fig:simplified_4x4_top}. Each of these blocks which make up the chiplet has an individual power profile. This means that different power profiles can be applied to every computational tile and router port in each chiplet. In our experimentation, different levels of detail are applied to the chiplets in the 2.5D and 3D systems as described in the following sections. 

\bh{2.5D System Specifics:} 
The target 2.5D system consists of a grid of chiplets integrated on an interposer, as illustrated in \rev{Fig.}~\ref{fig:simplified_4x4_top}. 
% Each chiplet consists of multiple layout blocks and router ports used for inter-chiplet communication, as detailed in Figure~\ref{fig:simplified_4x4_top}. Each of these individual routers and tiles is capable of an independent power profile to simulate varying communication or computation. 
Each chiplet is connected directly to the interposer via $\mu$-bumps surrounded by a capillary underfill material.
The physical dimensions of the router ports are compatible with Universal Chiplet Interconnect Express (UCIe) specification~\cite{sharma2022universal}. 
The entire package is covered by a copper lid, which contacts each chiplet through a thermal interface material (TIM).

\bh{3D System Specifics:}
In the 3D system, three stacked chiplets are placed in a 4x4 grid with equal spacing, consistent with~\cite{intel2006dieStacking}. 
Vertically stacked chiplets are connected to one another with $\mu$-bumps surrounded by a capillary underfill material. The bottom chiplet of each stack is connected to the interposer with the same method. This integration is illustrated in \rev{Fig.} ~\ref{fig:simplified_4x4_bottom}. The lid contacts only the top chiplet layer through a thermal interface material.
% The bottom chiplets are connected to the interposer through the $\mu$-bumps surrounded by a capillary underfill material, as detailed in Figure~\ref{fig:simplified_4x4_bottom}. The lid contacts only the top chiplet layer through a thermal interface material.

\subsection{Thermal RC Model Configuration} 
The number of nodes in the thermal RC network determines the model complexity, runtime, and granularity at which temperature can be observed in the model. 
A higher node density is used in chiplets to optimize this trade-off, while fewer nodes are used in non-chiplet components, such as the interposer, lid, substrate, and so on. 
%Without loss of generality, each chiplet in our experiments has four uniformly placed nodes (i.e., we monitor the temperature at each of these locations). 
Each chiplet is divided into four equal quadrants. One node is placed within each quadrant to allow for granular temperature monitoring across each chiplet. An alternate node density is used for non-chiplet layers in each model, as described below. This easily configurable non-uniform node density enables higher thermal resolution in critical parts such as chiplets while decreasing the runtime with lower resolution in less critical structures such as the substrate and lid. 
The DSS models in our experimentation are created by discretizing the thermal RC models with $T_s = 0.01 s$ sampling period. The sampling time can be chosen according to the application requirements.

\bh{2.5D Thermal RC Model Specifics:}
For the 2.5D systems, the choice of 4 nodes per chiplet leads to 64, 144, and 256 nodes in the chiplet layer of the 16, 36, and 64 chiplet systems, respectively.
For all other layers, the number of nodes is equal to the total number of chiplets per layer.
This allows the model to maintain higher thermal resolution in the critical chiplet layers while maintaining a fast execution time. 
% (16 for the 48 chiplet system since there are 16 chiplets per vertical layer).

\bh{3D Thermal RC Model Specifics:}
The node densities in the 3D system are adjusted similarly to the 2.5D models. The layers that contain chiplets use an 8$\times$8 grid, implying 4 nodes per chiplet. All other layers have a 4$\times$4 grid, leading to a lower node density.
The entire 3D system consists of 48 chiplets in total. 
There are a total of 192 nodes counting all nodes within chiplets. 
% 352 nodes total in 3D system

\subsubsection{Input Workloads and Power Consumption}
Identical workloads are considered for the 2.5D and 3D systems, with differences in chiplet power density detailed in the following subsections. The target chiplet systems are analyzed under one synthetic (WL1) and five real AI/ML application workloads (WL2-WL6).
The synthetic workload starts with a stress test that applies the maximum power to all chiplets to increase temperature beyond 100$^{\circ}$C. Then, a pseudo-random bit sequence (PRBS) is applied to each chiplet to emulate a wide range of dynamic variations. Finally, all chiplets are turned off to let the temperature return to the ambient state, as depicted in \rev{Fig.}~\ref{fig:temperature_vs_time_grid}.
Besides testing transient and steady-state behaviors, this power profile helps us to tune the thermal RC model.

% Table containing the list of and description of workloads
\begin{table}[b]
\renewcommand{\arraystretch}{0.9}
%\vspace{-12pt}
\caption{Descriptions of the workloads used in this work. (C) and (I) denote the CIFAR100 and ImageNet datasets, respectively.}
%\vspace{-8pt}
\label{tab:workloads}
% \resizebox{0.9\columnwidth}{!}{
\begin{tabular}{@{}lp{12cm}@{}}
\toprule
Workload& Composition \\ \midrule
WL1      & \rev{Synthetic (see Fig.~\ref{fig:temperature_vs_time_grid})} \\
WL2      & 16$\times$ResNet34 (C), 1$\times$VGG19 (C), 5$\times$ResNet50 (C), 3$\times$DenseNet40 (C), 1$\times$ResNet152 (C), 1$\times$VGG19 (I), 4$\times$ ResNet34 (I), 1$\times$ResNet18 (I), 1$\times$ResNet50 (I), 1$\times$VGG16 (I) \\
WL3      & 16$\times$ResNet34 (I), 1$\times$VGG19 (I), 5$\times$ResNet50 (I), 3$\times$DenseNet169 (I), 1$\times$ResNet110, 1$\times$VGG19 (I), 4$\times$ResNet101 (I), 1xResNet152 (I), 1$\times$ResNet18 (I), 1$\times$ResNet50 (I), 1$\times$Resnet152 (I) \\
WL4      & 16$\times$ResNet34 (C), 2$\times$VGG19 (I), 4$\times$DenseNet169 (I), 3$\times$DenseNet40 (C), 5$\times$ResNet50 (C), 3$\times$ResNet101, 7$\times$ResNet150 (I), 2$\times$VGG19 (I), 4$\times$ResNet101, 1$\times$VGG19 (C) \\
WL5      & 16$\times$Resnet34 (I), 1$\times$ResNet152 (I), 1$\times$ResNet110 (I), 3$\times$ResNet101 (I), 9$\times$DenseNet169 (I), 4$\times$ResNet34 (I), 12$\times$ResNet18 (I), 5$\times$ResNet50 (I), 1$\times$ResNet152 (I) \\
WL6      & 3$\times$DenseNet169 (I), 4$\times$ResNet34 (I), 12$\times$ResNet18 (I), 4$\times$ResNet101 (I), 2$\times$VGG19 (I), 4$\times$ResNet101 (I), 1$\times$VGG19 (C), 3$\times$DenseNet40 (C) \\ \bottomrule
\end{tabular}
% }
\end{table}

The remaining scenarios consider processing-in-memory (PIM)-based chiplets for accelerating ML workloads. The computational platform is resistive random access memory (ReRAM) based chiplets commonly used in literature~\cite{sharma2022swap, simba, song2017pipelayer}. We select this configuration due to its ability to efficiently implement matrix-vector multiplication, which is the predominant operation in any CNN workload. 
% Each chiplet consists of computational tiles along with peripheral circuits, such as accumulator, buffer, activation units, and pooling unit \cite{sharma2022swap, krishnan2021siam}. The tiles are connected through a network-on-chip (NoC). The hardware configuration within each chiplet is consistent with prior work \cite{sharma2023florets}. 
Each workload consists of a series of deep neural networks (DNNs) which run in series on the system. 
The workloads are listed in Table~\ref{tab:workloads}.
The neural networks (NN) in these workloads consist of several networks such as ResNets, DenseNets, and VGG networks. For example, WL1 contains 16 ResNet34's, followed by one VGG19, then 5 ResNet50's, and so on. Each workload contains from 20 to 40 individual networks. Workloads are mapped to the system as computing resources become available, meaning a new NN is mapped to chiplets when it completes the execution of a previous NN. 
Consequently, these workloads consist of NNs ranging from small NNs like ResNet18, which can be mapped to a single chiplet, to larger NNs such as DenseNet169, which are spread across multiple chiplets.

\rev{
After mapping the neural network workloads to the target chiplet-based system using a nearest-neighbor strategy inspired by the Simba~\cite{simba}, we estimate power consumption in two parts:
computation and communication. 
Computation power is obtained using NeuroSim~\cite{neurosim}, while communication power is estimated using a custom version of BookSim that includes interconnect energy modeling~\cite{jiang2013detailed}. 
We periodically sample the energy consumption from both tools at 10 ms intervals and convert these into dynamic power profiles for each chiplet. 
While these tools are not natively designed for chiplet systems, we follow a methodology adapted from SIAM that enables accurate per-chiplet power estimation for thermal simulations~\cite{krishnan2021siam}.
Importantly, while we use this methodology for our experiments, \textit{MFIT only requires power traces as input,} which can also be obtained from other simulators or real on-chip counters.
}

% After mapping the group of NNs to the target chiplet-based systems,  
% the chiplet power consumption is estimated in two parts, communication and computation. We estimate computation power through NeuroSim and interconnection network power using BookSim~\cite{neurosim, jiang2013detailed}. 
% We use running average power throughout the workload execution (40-55 seconds), consistent with power measuring tools such as Intel RAPL~\cite{david2010rapl} and pyNVML~\cite{pyNVML}. 

% New paragraph
\bh{Differences in 2.5D and 3D system chiplet power:}
Different hardware parameters, such as voltage and frequency, are used for the 2.5D and 3D systems. This results in lower per-chiplet power consumption in the 3D system of 1.2W as compared to 3W for the 2.5D system, detailed under the \textit{Power} section of Table ~\ref{tab:system_configurations}. Using these parameters, the total system power per lid area of the 3D system is between that of the 36 and 64 chiplet 2.5D system. 
This level means that the temperature of individual chiplets should be roughly equivalent between these systems, which is confirmed in \rev{Fig.}~\ref{fig:temperature_vs_time_grid}.

\begin{comment}
\todo{\subsubsection{HotSpot~\cite{skadron_temperatureAware} configuration:}
We also compare our proposed models to the state-of-the-art tool HotSpot~\cite{skadron_temperatureAware} for all the system sizes and workload configurations.
% The sampling period remained consistent at 0.1 seconds to match the thermal RC and FEM model settings. 
Since HotSpot was originally designed for 2D chips, we utilize an extension that adds 3D modeling capabilities~\cite{meng2012optimizing} to model both 2.5D and 3D integrated systems.
Geometry and material parameters are set to be identical to our reference FEM model. 
Since HotSpot does not support thermal conductivity variations in x-y-z directions, we use the average conductivity for anisotropic material layers (ex. C4 bump layer) in the chiplet package.
HotSpot also lacks the support for varying grid sizes across different layers. Therefore, we maintain a uniform grid size matching our chiplet layer.} 
%(64, 144, and 256 nodes for 16, 36, and 64 chiplet systems).
\end{comment}

\subsubsection{Configuration of other thermal simulators:}
We compare our proposed thermal RC and DSS models against the state-of-the-art analytical thermal modeling tools HotSpot~\cite{skadron_temperatureAware}, 3D-ICE~\cite{3D_ICE_3_1}, and PACT~\cite{pact} for all system sizes and workload configurations. 
Since HotSpot was originally designed for 2D chips, we utilize an extension that incorporates 3D modeling capabilities~\cite{meng2012optimizing} to support both 2.5D and 3D integrated systems. 
Identical geometry and material parameters are set across all tools, aligning them with our reference FEM model. 
None of these tools support thermal conductivity variations along the x, y, and z directions. Consequently, for anisotropic material layers (e.g., the C4 bump layer) in the chiplet package, we use an averaged thermal conductivity to approximate this configuration. 
Additionally, HotSpot and PACT do not allow non-uniform grid sizes across different horizontal layers, so a uniform grid size matching our chiplet layer is used. 
We configure the non-uniform grid in 3D-ICE~\cite{3D_ICE_3_1} to be consistent with our thermal RC model, ensuring the same number of thermal nodes for each system. 
Additionally, PACT provides flexibility in solver selection by supporting various SPICE solvers from Xyce~\cite{hutchinson2002xyce}. For consistency, we use the TRAP solver, as given in~\cite{pact}. 
PACT also enables parallel execution, which we leverage by running all simulations with eight cores on an Intel 10700k CPU.

% This section will show the performance (runtime) of the RC and SS model against FEM
%\vspace{-3mm}
\subsection{Execution Time Evaluation}\label{ssec:exe_time}
%\vspace{-0.5mm}
This section evaluates the execution time of the proposed multi-fidelity thermal model set. All simulations are run on a dual Intel Xeon Gold 6242R system with 40 processing cores. We use WL1 for our timing analysis since the execution time is comparable across all workloads.

\begin{figure*}[t]
    \centering
    \includegraphics[width=\textwidth]{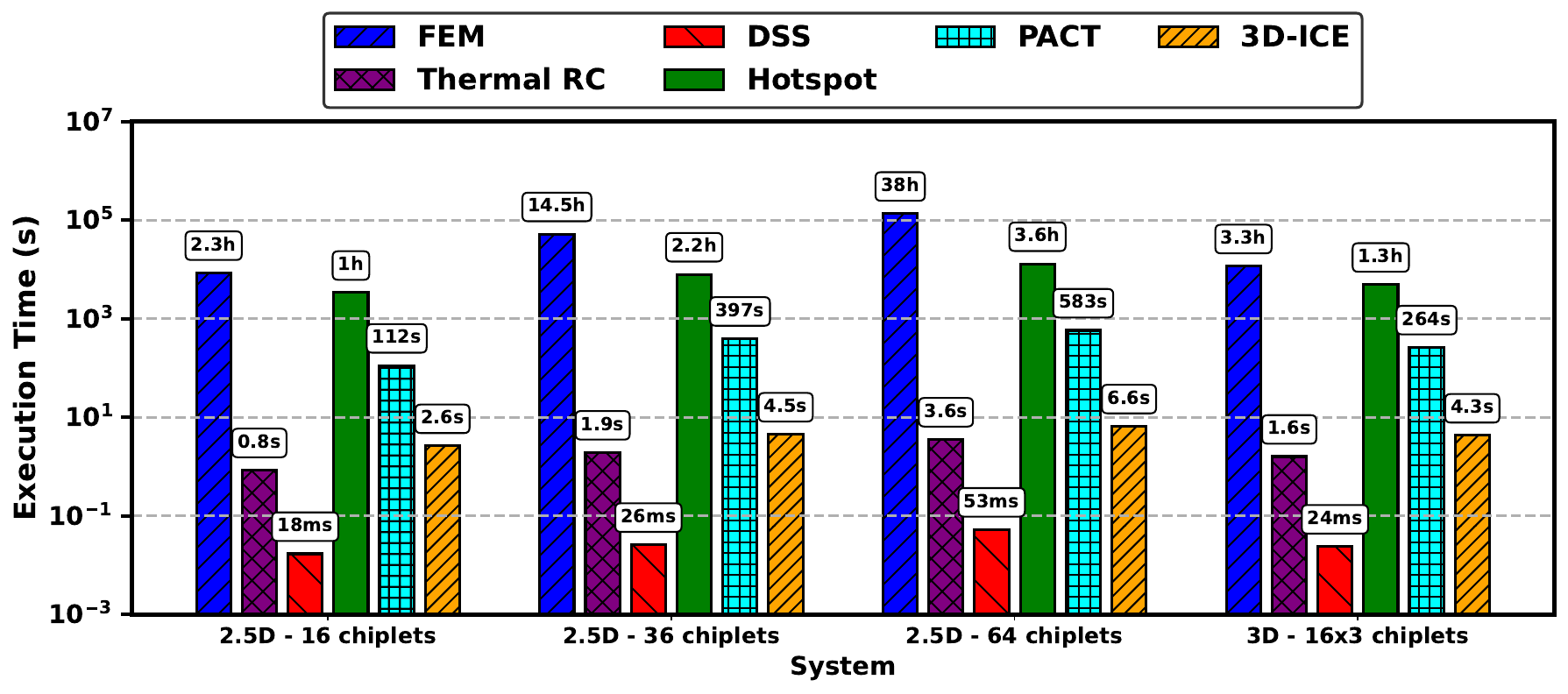}  % Adjust the width as needed
    \caption{Execution times (in log scale) of the proposed thermal models and HotSpot~\cite{skadron_temperatureAware}, PACT \cite{pact}, and 3D-ICE \cite{3D_ICE} for 16, 36, 64 - 2.5D integrated and 16x3 - 3D integrated chiplet systems.}
    \label{fig:exe_time}
\end{figure*}

\bh{2.5D Evaluation:}
Abstracted FEM simulations take 2.4, 14.5, and 38.0 hours for 16, 36, and 64 chiplet systems, respectively. 
While providing an accurate reference, these long simulation times and significant development effort make FEM impractical for DSE. 
Our thermal RC models fill this gap with execution times ranging from 0.85 to 3.57 seconds for 16, 36, and 64 chiplets, as summarized in the first 3 systems in \rev{Fig.}~\ref{fig:exe_time}. 
Coupled with the accuracy presented in the Section~\ref{ssec:accuracy}, the 10,000 to 40,000-fold speedup demonstrates their viability as a DSE tool.

Thermal RC models are derived directly from the underlying geometry and material parameters, meaning they can be reconfigured for different hardware and design configurations without re-calibration from FEM simulations. 
Relaxing this physical system to model connection, our DSS models reduce execution time to 18, 26, and 54 milliseconds for 16, 36, and 64 chiplets respectively. 
This speedup with respect to the RC model enables runtime temperature prediction, which can inform dynamic thermal power management (DTPM) decisions to increase system performance and reliability \cite{brooks2007thermalReliability}.
\rev{Although accurate real-time power estimation is a critical component of such DTPM frameworks, MFIT treats power as an external input and can be used with power traces obtained from simulators, profilers, or real hardware counters. 
For instance, physical current sensors that measure runtime power consumption at cluster or core granularities can provide direct input to MFIT~\cite{cochet2016chip, chan2013chip}.
Another option is software-based power meters that estimate runtime power by converting performance counter readings into power values, such as Qualcomm's profiler~\cite{qualcomm_profiler}.}
While maintaining the accuracy for a given configuration, DSS can be regenerated from the RC model in a few milliseconds if the sampling period or hardware configuration changes. 

\bh{3D Evaluation:}
The execution time results of the 3D system are summarized in the 3D - 16x3 chiplet system of \rev{Fig.}~\ref{fig:exe_time}. FEM simulations take approximately 3.3 hours for the single 3D system. 
% The execution time of the 3D FEM model is faster due to the simplified geometry used compared to the 2.5D models. In the 2.5D models, 12 individual bodies were used to model each chiplet, whereas in the 3D model only one body per chiplet is used. 
The thermal RC model significantly reduces the runtime to 1.6 seconds. Similar to the 2.5D system, the 3D DSS model further reduces runtime to just 24 milliseconds.
This substantial speedup is due to the DSS model relying solely on matrix multiplication operations, which are highly optimized on modern hardware through SIMD instructions and other acceleration techniques, such as efficient memory access.
%For comparison, the execution time of a similarly sized 2.5D system DSS model is 0.09 seconds. 

% Origin Execution time figure
% \begin{figure}[t]
% \centering
%     \centering
%     \includegraphics{figures/execution_time/exe_time_log.pdf}
%     % \Description{...}
%     % \vskip -10pt
%     \caption{Execution times (in log scale) of our thermal models and HotSpot~\cite{skadron_temperatureAware} for 16, 36, 64 - 2.5D integrated and 16x3 - 3D integrated chiplet systems.}
%     \label{fig:exe_time}
%     % \vskip -15pt
% \end{figure}

\bh{Execution time comparison to existing tools:} 
The commonly used thermal modeling tools, HotSpot~\cite{hotspot}, 3D-ICE~\cite{3D_ICE_3_1}, and PACT~\cite{pact} belong to the same class of model as our thermal RC models. As shown in \rev{Fig.}~\ref{fig:exe_time}, our proposed thermal RC and DSS methods achieve significantly faster execution times compared to these tools.

\rev{HotSpot and PACT lack support for non-uniform grid structures, resulting in more complex ODE formulations and increased computational cost. 
In particular, HotSpot relies on the classical Runge-Kutta 4 (RK4) method, which is known to be less efficient for stiff systems due to its small time step requirements to maintain stability and accuracy, leading to increased computation time.}
As a result, our thermal RC model achieves a 4235$\times$ speedup for 16 chiplets, 4082$\times$ for 36 chiplets, 3630$\times$ for 64 chiplets, and 4050$\times$ for the 16$\times$3 3D chiplet system compared to HotSpot. 
Against PACT, our thermal RC model is 132$\times$ faster for 16 chiplets, 205$\times$ for 36 chiplets, 164$\times$ for 64 chiplets, and 167$\times$ for the 16$\times$3 3D chiplet system.

\rev{While 3D-ICE does support non-uniform grids and employs an implicit Euler method for solving transient heat equations, MFIT further improves efficiency by integrating BLAS-optimized routines within the ODE solver pipeline. These routines accelerate matrix-vector multiplications and factorization steps, enabling faster convergence when used alongside the SuperLU solver. In contrast, 3D-ICE relies solely on SuperLU without these enhancements.}
Consequently, our approach achieves a 3$\times$ speedup for 16 chiplets, 2.3$\times$ for 36 chiplets, 1.9$\times$ for 64 chiplets, and 2.7$\times$ for the 16$\times$3 3D chiplet system over 3D-ICE.
Moreover, our DSS method surpasses all RC-based models in execution efficiency, demonstrating the best performance among all evaluated tools. 
Finally, we emphasize that MFIT, our multi-fidelity thermal model set, covers a \textit{much wider range of accuracy and execution time trade-offs} than a specific point solution, providing greater flexibility for different modeling needs.

\begin{comment}
\todo{
\bh{Execution time comparison to HotSpot~\cite{skadron_temperatureAware}:} 
The commonly used thermal modeling tool HotSpot belongs to the same class as our thermal RC models. 
The execution times of the proposed models are significantly faster (1862$\times$ for 16, 607$\times$ for 36, and 245$\times$ for 64 - 2.5D integrated chiplet systems while 817$\times$ for the 16x3 - 3D integrated chiplet system) than HotSpot, as shown in Figure~\ref{fig:exe_time}.
The significant speedup mentioned above with similar or higher accuracy is primarily attributed to two factors. 
First, we employ a non-uniform grid for different layers, as explained in Section~\ref{ssec:FEM_to_RC}. 
The speedup due to the use of a non-uniform grid size is 
13$\times$, 17$\times$, 19$\times$ and 7$\times$ faster execution for 16, 36, 64 - 2.5D integrated and 16x3 - 3D integrated chiplet systems with respect to HotSpot.
Second, we employ the adaptive solver LSODA. This solver requires fewer iterations per time step for convergence in comparison to HotSpot. Utilizing an identical grid configuration to HotSpot, the use of this solver alone leads to speedups of approximately 144$\times$, 35$\times$, 13$\times$, and 122$\times$, again with respect to the baseline HotSpot model execution time.
Finally, we emphasize that MFIT, our multi-fidelity thermal model set covers a much wider range of accuracy and execution time trade-offs than a specific point solution such as HotSpot.
}
\end{comment}

% This section will show the temperature accuracy validation of RC and SS against FEM
\subsection{Validation of Thermal RC and DSS model}
\label{ssec:accuracy}
% Adjusted 2x2 grid of Temperature vs Time plots with increased vertical spacing
\begin{figure*}[b]  % Use figure* for two-column formats
    \centering
    % First Row of Subfigures
    \begin{subfigure}[b]{0.4\textwidth}
        \includegraphics[width=\linewidth]{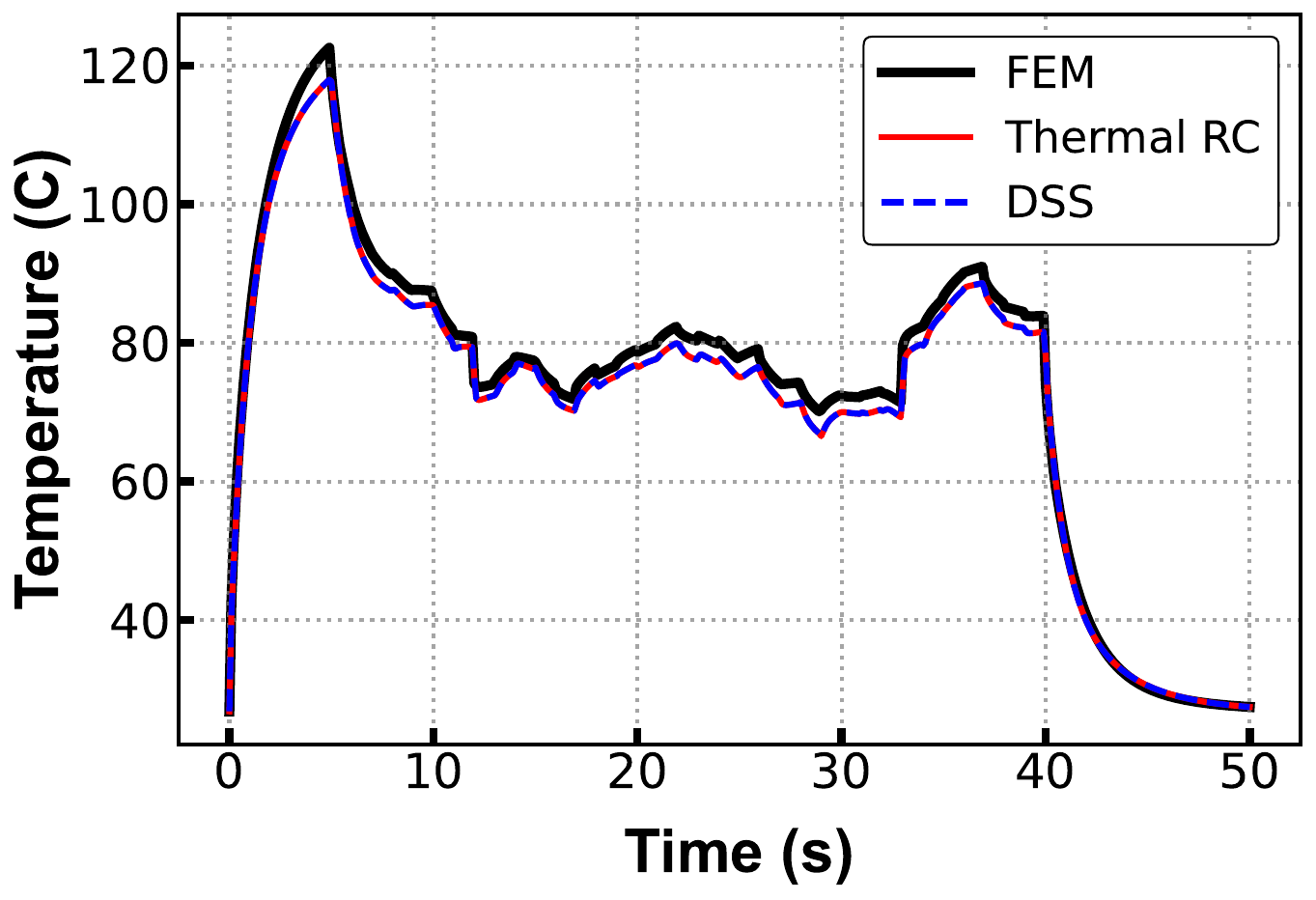}
        \caption{2.5D - 16 chiplet system}
        \label{fig:temperature_vs_time_2.5d_16}
    \end{subfigure}
    \hspace{30pt}
    \begin{subfigure}[b]{0.4\textwidth}
        \includegraphics[width=\linewidth]{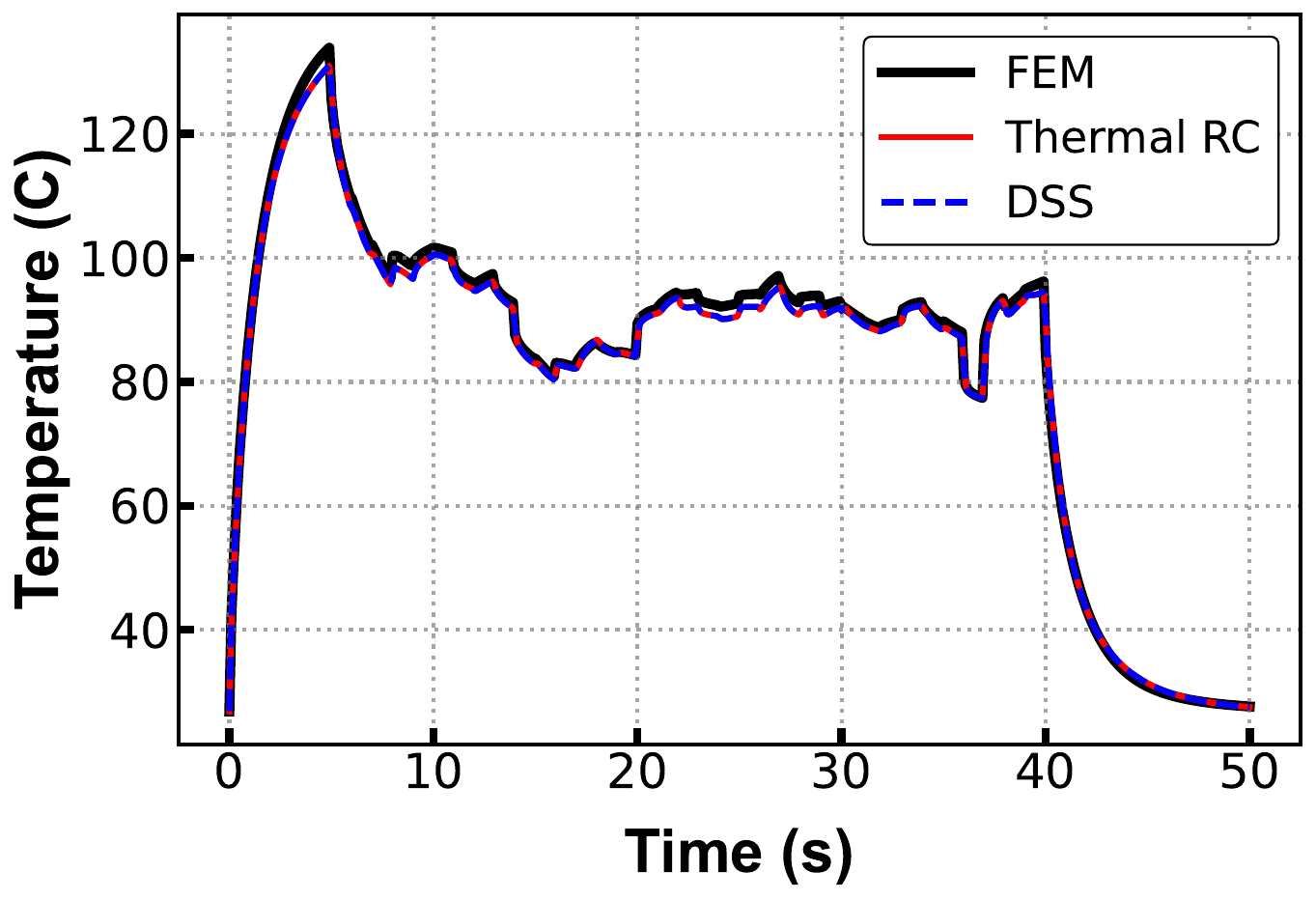}
        \caption{2.5D - 36 chiplet system}
        \label{fig:temperature_vs_time_2.5d_36}
    \end{subfigure}

    % Add vertical space between rows
    \vspace{10pt}

    % Second Row of Subfigures
    \begin{subfigure}[b]{0.4\textwidth}
        \includegraphics[width=\linewidth]{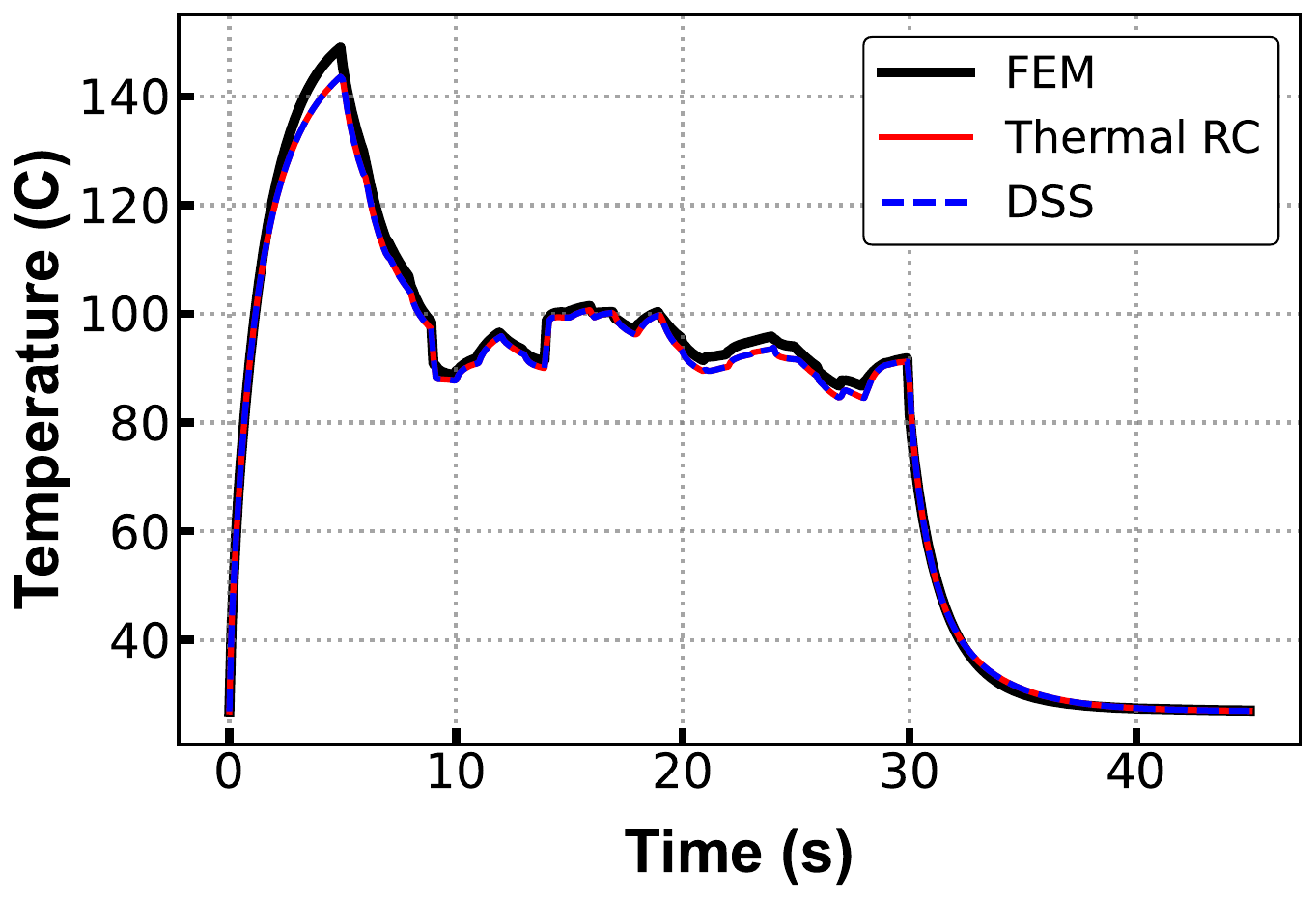}
        \caption{2.5D - 64 chiplet system}
        \label{fig:temperature_vs_time_2.5d_64}
    \end{subfigure}
    \hspace{30pt}
    \begin{subfigure}[b]{0.4\textwidth}
        \includegraphics[width=\linewidth]{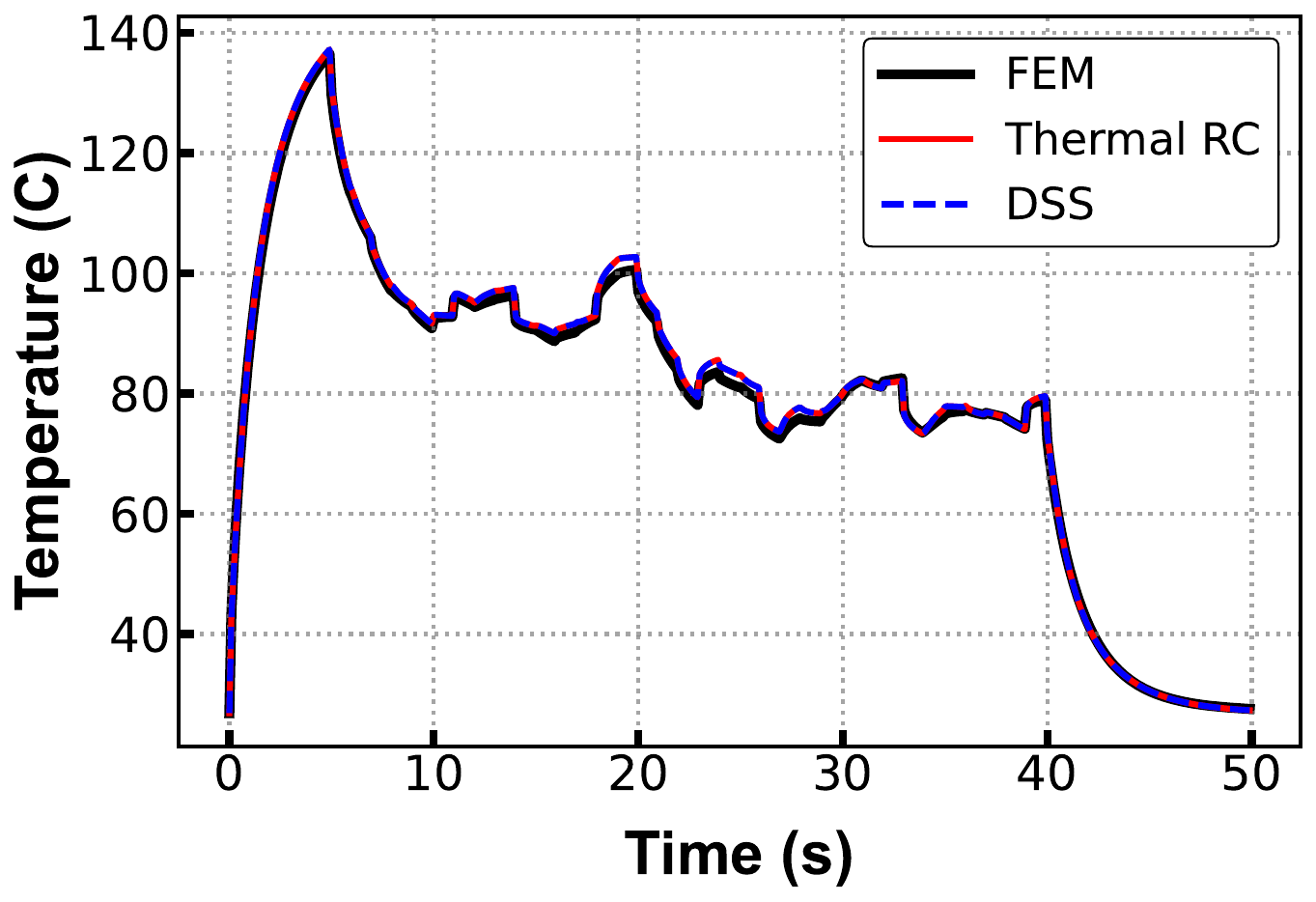}
        \caption{3D - 16x3 chiplet system}
        \label{fig:temperature_vs_time_3d_48}
    \end{subfigure}

    % Main caption for the entire figure
    \caption{\rev{Temperatures of representative chiplets from 2.5D and 3D systems while running WL1 are plotted as a function of time.}}
    \label{fig:temperature_vs_time_grid}
\end{figure*}

We validate the accuracy of our thermal RC and DSS models by comparing their temperature estimates to full-system FEM simulation results for the same workload and system configurations. This comparison is completed for each of the three 2.5D system sizes and the single 3D system. 
In addition to the visualization of the temperature estimate over time, shown in \rev{Fig.}~\ref{fig:temperature_vs_time_grid}, 
\rev{three metrics are used to quantify the accuracy of our thermal RC and DSS models against the FEM results. The MAE metric measures the mean absolute error in temperature across the entire simulation duration. The second metric, average percentage error, expresses the temperature error as a percentage. 
This is particularly important because a given absolute error at lower temperatures represents a more significant inaccuracy than the same absolute error at higher temperatures.
% This is particularly important because a given absolute error at lower temperatures poses a greater risk than the same error at higher temperatures. 
The third metric,}
Temperature Violation Prediction accuracy, measures the accuracy of our models in predicting temperature violations. Predicting temperature violations (e.g., tracking the time steps when the temperature exceeds the allowed threshold) is often used by DTPM algorithms. We set 85$^{\circ}$C as the maximum allowable temperature threshold for each system without loss of generality~\cite{zhou2019temperature}. 
This metric first identifies the time steps in FEM simulations where temperature violations occur (temperature exceeds 85$^{\circ}$C).
Then, it computes the percentage of these violations captured by the thermal RC and DSS models (e.g., 100\% means all violations are detected with perfect accuracy). The proposed models conservatively flag violations within one degree of the above-mentioned threshold temperature.

\begin{figure*}[b]
    \centering
    \begin{subfigure}[b]{0.49\textwidth}
        \includegraphics[width=\columnwidth]{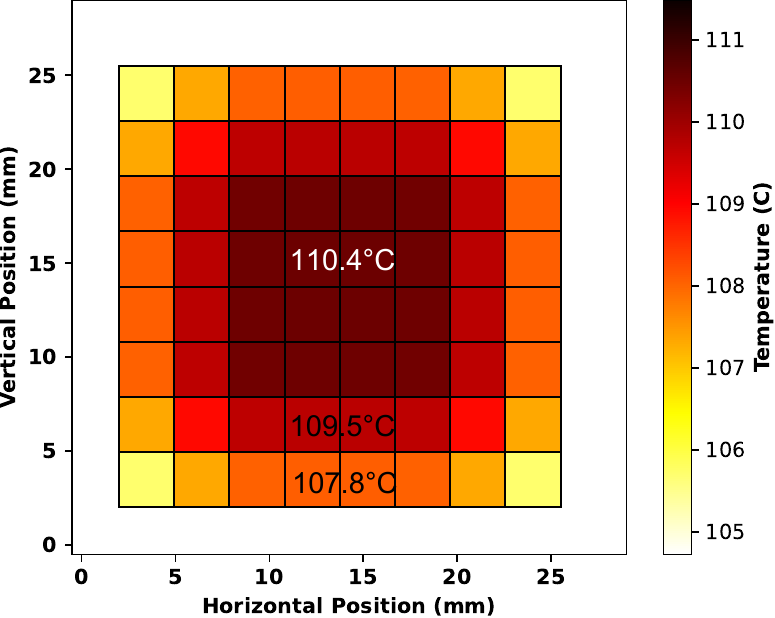}
        \caption{}
    \end{subfigure}
    \begin{subfigure}[b]{0.49\textwidth}
        \includegraphics[width=\linewidth]{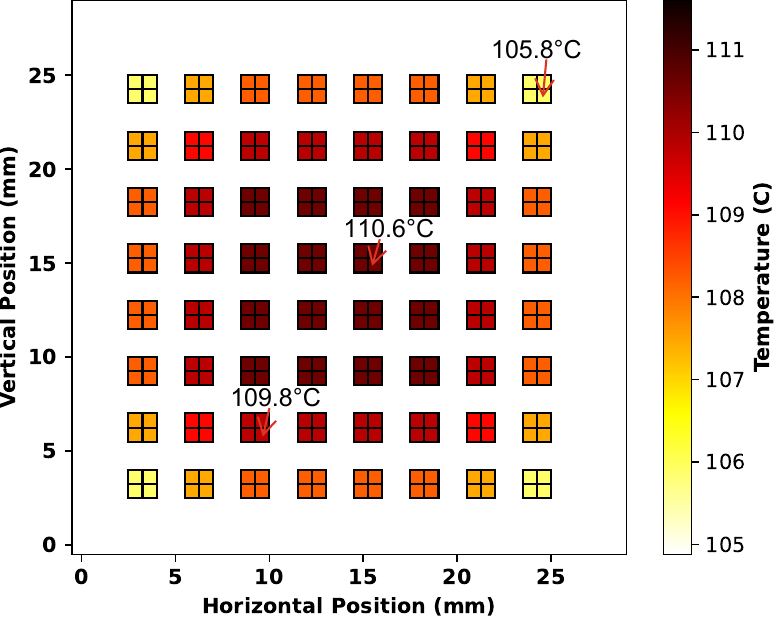}
        \caption{}
    \end{subfigure}
    \caption{\rev{The heat map of (a) the interposer layer and (b) chiplets generated by the thermal RC model for 64 chiplet system.}}
    \label{fig:rc_heatmap}
\end{figure*}

To assist users in visualizing the thermal behavior of the system under test, the RC model also creates a heat map of each layer in the system. As an example, the heat map of the interposer layer of a 2.5D 64 chiplet system is shown in \rev{Fig.}~\ref{fig:rc_heatmap}. This figure shows the temperature gradient that occurs between the center of the interposer, where the heat-producing chiplets are located, and the edges of the system, where there are no chiplets. These maps allow for quick visual verification of the system behavior instead of relying purely on numerical results.

% RC Model temperature contour

% % 3D chiplet plot
% \begin{figure}[t]
% \centering
%     \centering
%     \includegraphics[width=0.9\columnwidth]{figures/accuracy_verification_figures/3D_Chiplet_Temperature_Comparison.pdf}
%     % \Description{...}
%     % \vskip -10pt
%     \caption{Temperature vs. Time of a representative chiplet from the single 3D system. WL1 is shown.}
%     \label{fig:3D_temperature_comparison}
%     % \vskip -15pt
% \end{figure}

\bh{2.5D Validation Results:}
\rev{Figs.}~\ref{fig:temperature_vs_time_2.5d_16}, \ref{fig:temperature_vs_time_2.5d_36}, and \ref{fig:temperature_vs_time_2.5d_64} plot the temperature as a function of time for each 2.5D system size of a representative chiplet while running workload WL1.
All three plots of 2.5D systems clearly show that the systematically constructed thermal RC and DSS models produce near-identical results to the FEM baseline. 
They closely follow the FEM results during the stress test (temperature increases until reaching the maximum point), randomly changing chiplet power consumption (middle portion), and cool-down periods. 

% Adjusted 2x2 grid of Temperature vs Time plots with increased vertical spacing

% Table with accuracy results for 2.5D experiments
%\input{files/accuracy_table_2.5D}

% Table with accuracy results for 3D experiments
%\input{files/accuracy_table_3D}

% Table with the results of all experiments
\begin{table*}[ht]
    \caption{\rev{Comparison of thermal RC, DSS, HotSpot ~\cite{skadron_temperatureAware}, 3D-ICE~\cite{3D_ICE}, and PACT~\cite{pact} models based on mean absolute error, average percentage error, and accuracy in predicting temperature violations relative to the reference FEM models. The best MAE and percentage error for each system-workload combination are highlighted in red. \textbf{Average operating temperatures are in the 60-80 ($\circ$C) range.}}}
    \centering
    \renewcommand{\arraystretch}{0.98}
    \setlength{\tabcolsep}{2pt} % Adjust column separation
    \scriptsize % Reduce font size to fit the table on the page

    % Begin the tabular environment with appropriate column definitions
    \begin{tabular}{@{}ll|ccc|ccc|ccc|ccc@{}}
        \toprule
        % First row: Headers for each section
        \multicolumn{2}{c|}{} & \multicolumn{3}{c|}{\textbf{2.5D - 16 Chiplets}} & \multicolumn{3}{c|}{\textbf{2.5D - 36 Chiplets}} & \multicolumn{3}{c|}{\textbf{2.5D - 64 Chiplets}} & \multicolumn{3}{c}{\textbf{3D - 16x3 Chiplets}} \\ \midrule

        % Second row: Sub-headers for each section
        \textbf{WL} & \textbf{Model} & \textbf{MAE} & \textbf{\begin{tabular}[c]{@{}c@{}}\rev{Avg.}\\\rev{Error}\end{tabular}} & \textbf{\begin{tabular}[c]{@{}c@{}}Temp. Violation\\ Accuracy\end{tabular}} &
        \textbf{MAE} & \textbf{\begin{tabular}[c]{@{}c@{}}\rev{Avg.}\\\rev{Error}\end{tabular}} & \textbf{\begin{tabular}[c]{@{}c@{}}Temp. Violation\\ Accuracy\end{tabular}} &
        \textbf{MAE} & \textbf{\begin{tabular}[c]{@{}c@{}}\rev{Avg.}\\\rev{Error}\end{tabular}} & \textbf{\begin{tabular}[c]{@{}c@{}}Temp. Violation\\ Accuracy\end{tabular}} &
        \textbf{MAE} & \textbf{\begin{tabular}[c]{@{}c@{}}\rev{Avg.}\\\rev{Error}\end{tabular}} & \textbf{\begin{tabular}[c]{@{}c@{}}Temp. Violation\\ Accuracy\end{tabular}} \\ \midrule
        \multirow{5}{*}{WL1} & Thermal RC & \highlight{1.23$^\circ$C} & \highlight{1.59\%} & 93.5\% & \highlight{1.42$^\circ$C} & \highlight{1.62\%} & 95.7\% & \highlight{0.99$^\circ$C} & \highlight{1.26\%} & 99.8\% & \highlight{0.94$^\circ$C} & \highlight{1.11\%} & 98.1\% \\
        & DSS & \highlight{1.23$^\circ$C} & \highlight{1.59\%} & 93.5\% & \highlight{1.42$^\circ$C} & \highlight{1.62\%} & 95.7\% & \highlight{0.99$^\circ$C} & \highlight{1.26\%} & 99.8\% & \highlight{0.94$^\circ$C} & \highlight{1.11\%} & 98.1\% \\
        & \textit{HotSpot} & 2.74$^\circ$C & \rev{3.78\%} & 67.8\% & 1.64$^\circ$C & \rev{2.13\%} & 95.2\% & 2.16$^\circ$C & \rev{2.62\%} & 95.7\% & 2.37$^\circ$C & \rev{2.99\%} & 87.5\% \\
        & \textit{3D-ICE} & 1.65$^\circ$C & \rev{2.28\%} & 89.5\% & 2.07$^\circ$C & \rev{2.63\%} & 97.4\% & 1.65$^\circ$C & \rev{2.07\%} & 99.5\% & 1.73$^\circ$C & \rev{2.21\%} & 96.9\% \\
        & \textit{PACT} & 1.99$^\circ$C & \rev{2.76\%} & 91.7\% & 1.61$^\circ$C & \rev{2.13\%} & 99.1\% & 1.70$^\circ$C & \rev{2.15\%} & 99.4\% & 2.09$^\circ$C & \rev{2.62\%} & 98.7\% \\
        \midrule
        \multirow{5}{*}{WL2} & Thermal RC & \highlight{0.86$^\circ$C} & \highlight{1.38\%} & 96.0\% & 1.16$^\circ$C & \rev{1.89\%} & 0.0\% & \highlight{0.95$^\circ$C} & \highlight{1.56\%} & 100.0\% & \highlight{0.98$^\circ$C} & \highlight{1.58\%} & 100.0\% \\
        & DSS & \highlight{0.86$^\circ$C} & \highlight{1.38\%} & 96.0\% & 1.16$^\circ$C & \rev{1.89\%} & 0.0\% & \highlight{0.95$^\circ$C} & \highlight{1.56\%} & 100.0\% & \highlight{0.98$^\circ$C} & \highlight{1.58\%} & 100.0\% \\
        & \textit{HotSpot} & 1.58$^\circ$C & \rev{2.80\%} & 75.0\% & 1.24$^\circ$C & \rev{1.95\%} & 0.0\% & 1.01$^\circ$C & \rev{1.62\%} & 100.0\% & 1.74$^\circ$C & \rev{2.70\%} & 100.0\% \\
        & \textit{3D-ICE} & 1.36$^\circ$C & \rev{2.21\%} & 85.1\% & 1.35$^\circ$C & \rev{2.13\%} & 0.0\% & 1.82$^\circ$C & \rev{2.93\%} & 100.0\% & 1.39$^\circ$C & \rev{2.21\%} & 100.0\% \\
        & \textit{PACT} & 1.58$^\circ$C & \rev{2.38\%} & 75.9\% & \highlight{0.91$^\circ$C} & \highlight{1.44\%} & 16.7\% & 1.55$^\circ$C & \rev{2.51\%} & 100.0\% & 2.10$^\circ$C & \rev{3.33\%} & 100.0\% \\
        \midrule
        \multirow{5}{*}{WL3} & Thermal RC & \highlight{1.02$^\circ$C} & \highlight{1.63\%} & 100.0\% & 1.28$^\circ$C & \rev{1.93\%} & 78.5\% & \highlight{1.03$^\circ$C} & \highlight{1.57\%} & 95.8\% & \highlight{0.80$^\circ$C} & \highlight{1.23\%} & 100.0\% \\
        & DSS & \highlight{1.02$^\circ$C} & \highlight{1.63\%} & 100.0\% & 1.28$^\circ$C & \rev{1.93\%} & 78.5\% & \highlight{1.03$^\circ$C} & \highlight{1.57\%} & 95.8\% & \highlight{0.80$^\circ$C} & \highlight{1.23\%} & 100.0\% \\
        & \textit{HotSpot} & 1.97$^\circ$C & \rev{3.02\%} & 100.0\% & 1.18$^\circ$C & \rev{1.74\%} & 29.7\% & 1.12$^\circ$C & \rev{1.65\%} & 67.4\% & 1.79$^\circ$C & \rev{2.67\%} & 100.0\% \\
        & \textit{3D-ICE} & 1.15$^\circ$C & \rev{1.79\%} & 100.0\% & 1.53$^\circ$C & \rev{2.27\%} & 3.8\% & 1.82$^\circ$C & \rev{2.75\%} & 76.4\% & 1.12$^\circ$C & \rev{1.72\%} & 100.0\% \\
        & \textit{PACT} & 2.03$^\circ$C & \rev{3.12\%} & 100.0\% & \highlight{1.09$^\circ$C} & \highlight{1.61\%} & 23.4\% & 1.75$^\circ$C & \rev{2.59\%} & 66.7\% & 1.80$^\circ$C & \rev{2.76\%} & 100.0\% \\
        \midrule
        \multirow{5}{*}{WL4} & Thermal RC & \highlight{1.46$^\circ$C} & \highlight{1.95\%} & 96.6\% & \highlight{1.64$^\circ$C} & \highlight{2.10\%} & 96.0\% & \highlight{1.35$^\circ$C} & \highlight{1.74\%} & 98.0\% & \highlight{1.48$^\circ$C} & \highlight{1.85\%} & 98.0\% \\
        & DSS & \highlight{1.46$^\circ$C} & \highlight{1.95\%} & 96.6\% & \highlight{1.64$^\circ$C} & \highlight{2.10\%} & 96.0\% & \highlight{1.35$^\circ$C} & \highlight{1.74\%} & 98.0\% & \highlight{1.48$^\circ$C} & \highlight{1.85\%} & 98.0\% \\
        & \textit{HotSpot} & 2.29$^\circ$C & \rev{3.36\%} & 95.6\% & 7.39$^\circ$C & \rev{10.28\%} & 92.4\% & 2.10$^\circ$C & \rev{2.59\%} & 96.5\% & 2.71$^\circ$C & \rev{3.36\%} & 96.2\% \\
        & \textit{3D-ICE} & 3.72$^\circ$C & \rev{4.72\%} & 91.3\% & 2.33$^\circ$C & \rev{2.94\%} & 97.2\% & 2.68$^\circ$C & \rev{3.43\%} & 97.3\% & 2.46$^\circ$C & \rev{3.08\%} & 96.4\% \\
        & \textit{PACT} & 3.56$^\circ$C & \rev{4.74\%} & 93.8\% & 1.94$^\circ$C & \rev{2.46\%} & 97.6\% & 2.39$^\circ$C & \rev{3.06\%} & 97.2\% & 2.97$^\circ$C & \rev{3.79\%} & 96.3\% \\
        \midrule
        \multirow{5}{*}{WL5} & Thermal RC & \highlight{1.01$^\circ$C} & \highlight{1.62\%} & 100.0\% & 1.25$^\circ$C & \rev{1.89\%} & 78.4\% & \highlight{0.96$^\circ$C} & \highlight{1.43\%} & 100.0\% & 0.87$^\circ$C & \rev{1.35\%} & 100.0\% \\
        & DSS & \highlight{1.01$^\circ$C} & \highlight{1.62\%} & 100.0\% & 1.25$^\circ$C & \rev{1.89\%} & 78.4\% & \highlight{0.96$^\circ$C} & \highlight{1.43\%} & 100.0\% & 0.87$^\circ$C & \rev{1.35\%} & 100.0\% \\
        & \textit{HotSpot} & 1.95$^\circ$C & \rev{3.03\%} & 100.0\% & 1.16$^\circ$C & \rev{1.71\%} & 57.9\% & 1.30$^\circ$C & \rev{1.88\%} & 0.0\% & 1.86$^\circ$C & \rev{2.77\%} & 100.0\% \\
        & \textit{3D-ICE} & 1.16$^\circ$C & \rev{1.82\%} & 100.0\% & 1.53$^\circ$C & \rev{2.27\%} & 15.8\% & 1.79$^\circ$C & \rev{2.67\%} & 50.0\% & \highlight{0.87$^\circ$C} & \highlight{1.33\%} & 100.0\% \\
        & \textit{PACT} & 1.99$^\circ$C & \rev{3.09\%} & 100.0\% & \highlight{1.13$^\circ$C} & \highlight{1.68\%} & 38.4\% & 1.87$^\circ$C & \rev{2.72\%} & 15.0\% & 1.64$^\circ$C & \rev{2.51\%} & 100.0\% \\
        \midrule
        \multirow{5}{*}{WL6} & Thermal RC & \highlight{0.89$^\circ$C} & \highlight{1.44\%} & 98.1\% & 1.30$^\circ$C & \rev{1.97\%} & 84.3\% & \highlight{1.21$^\circ$C} & \highlight{1.79\%} & 89.7\% & \highlight{1.52$^\circ$C} & \highlight{2.36\%} & 91.2\% \\
        & DSS & \highlight{0.89$^\circ$C} & \highlight{1.44\%} & 98.1\% & 1.30$^\circ$C & \rev{1.97\%} & 84.3\% & \highlight{1.21$^\circ$C} & \highlight{1.79\%} & 89.7\% & \highlight{1.52$^\circ$C} & \highlight{2.36\%} & 91.2\% \\
        & \textit{HotSpot} & 1.62$^\circ$C & \rev{2.82\%} & 85.9\% & \highlight{1.28$^\circ$C} & \highlight{1.88\%} & 87.8\% & 1.48$^\circ$C & \rev{2.18\%} & 76.1\% & 2.22$^\circ$C & \rev{3.37\%} & 82.3\% \\
        & \textit{3D-ICE} & 1.34$^\circ$C & \rev{2.17\%} & 91.8\% & 1.69$^\circ$C & \rev{2.48\%} & 74.7\% & 2.38$^\circ$C & \rev{3.54\%} & 69.8\% & 1.61$^\circ$C & \rev{2.44\%} & 70.3\% \\
        & \textit{PACT} & 1.67$^\circ$C & \rev{2.56\%} & 89.3\% & 1.28$^\circ$C & \rev{1.91\%} & 91.4\% & 1.85$^\circ$C & \rev{2.75\%} & 90.2\% & 2.43$^\circ$C & \rev{3.69\%} & 66.9\% \\
        \bottomrule
    \end{tabular}
    \label{tab:combined_accuracy}
\end{table*}

The first 3 columns of Table~\ref{tab:combined_accuracy} summarize the accuracy results for all 2.5D systems and workloads.
\rev{. The worst-case mean absolute errors of our models are only 1.46$^{\circ}$C, 1.64$^{\circ}$C, and 1.35$^{\circ}$C for the 16-, 36-, and 64-chiplet systems, respectively. The corresponding worst-case average percentage errors are 1.95\%, 2.10\%, and 1.79\%, indicating that even the largest errors occur at higher temperatures where such deviations are less critical. In contrast, other tools exhibit significantly higher errors, as discussed later.} 
These results indicate that the proposed models achieve excellent accuracy across different hardware configurations and workloads.
Our models also achieve FEM-level high accuracy in predicting temperature violations as described earlier in Section \ref{ssec:accuracy}. For example, the worst-case accuracy for the 16-chiplet system is 93.5\% (i.e., only 6.5\% of the time steps where violations are missed). 
With the exception of WL2 and 3, prediction accuracy exceeds 82\% for all other workloads run on the 36 chiplet system. The same trend is seen with the 64-chiplet system.
% Our models capture 100\% of the violations during WL3 and WL5 while missing a handful of them for other workloads. 
% The corresponding accuracy for the 36-chiplet system is above 95\% for WL1 and WL4. 
% We observe 77.2\%, 84.8\%, and 87.8\% accuracy for WL3, WL6, and WL5, respectively. 

For some specific workload-system combinations, such as WL2 and 3 run on the 36 chiplet system, prediction accuracy falls below the 80\% mark for our models. 
These relatively lower accuracy values stem from sudden temperature spikes that lead to short-term temperature violations in these workload-system combinations. Temperature spikes in specific chiplets occur when several grouped chiplets experience a transient power spike simultaneously. This increases the temperature of the more central chiplets in the group. In these cases, the peak temperatures are mostly at or below 85 degrees with infrequent short-term violations, which can be tolerated.
For example, FEM simulations indicate only 255 temperature violations across all chiplets in WL3 compared to 11 thousand violations in WL1. Hence, missing even a few violations impacts the accuracy heavily when the RC and DSS models do not capture these spikes. 
In contrast, our thermal RC and DSS models effectively detect more prolonged violations, as evidenced by WL1, WL3, WL4, WL5, and WL6. 
% Similarly, our thermal RC and DSS models achieve high accuracy for the 64-chiplet system, as shown in Table~\ref{tab:combined_accuracy}. The lowest accuracies are 82.4\% for WL5 and 89.3\% for WL3, which have few total violations.

\bh{3D Validation Results:}
\rev{Fig.}~\ref{fig:temperature_vs_time_3d_48} plots the temperature as a function of time for the single 3D system of a representative chiplet running WL1. The plot shows similar behavior to the 2.5D comparison, where the RC and DSS match each other exactly and match the reference FEM results extremely closely during the stress test, random, and cool-down portions of the workload.

\rev{The far-right column of Table~\ref{tab:combined_accuracy} summarizes the accuracy results for the single 3D system for each workload. The worst-case mean absolute error is only 1.52$^{\circ}$C, while the worst-case average percentage error is 2.36\%.  This result shows that the proposed models maintain high levels of accuracy even when applied to a stacked-die 3D package.}
This high degree of accuracy is repeated when predicting temperature violations. The worst-case temperature violation prediction accuracy is only 94.0\%, occurring for workload 6.

\bh{Accuracy comparison to existing tools:}
Table~\ref{tab:combined_accuracy} also lists the MAE of HotSpot, 3D-ICE, and PACT simulations compared to FEM simulations for the 2.5D and 3D systems. For the 2.5D system results, the average error of Hotspot, 3D-ICE, and PACT across all workloads are 0.7, 0.66, and 0.59 degrees greater than our thermal RC and DSS models. 
For the 3D system results, the average error of Hotspot, 3D-ICE, and PACT across all workloads are 1.02, 0.43, and 1.07 degrees greater than our thermal RC and DSS models.

There are several reasons for the improved accuracy of our models as compared to existing thermal modeling tools. First, PACT and 3D-ICE are not able to model convective heat transfer from more than one boundary of the package. In the case of 3D-ICE, this configuration would require a uniform grid node distribution, which would negatively impact execution time. Second, none of these existing simulators support the ability to provide different thermal conductivities along the x, y, and z-axis, further decreasing accuracy. Third, our models have their node capacitance values tuned, improving their transient thermal simulation accuracy.
\rev{Lastly, the strength of the PACT simulator lies in circuit-level thermal simulations~\cite{pact}. It assumes a much finer grid granularity compared to block granularity, meaning that each block contains many thermal nodes. However, in architectural-level thermal simulations, the focus of this work, each block contains significantly fewer nodes. This mismatch can lead to uneven node distribution within the PACT simulator, potentially reducing simulation accuracy.}

\subsection{Modeling of Heterogeneous Architecture}
\begin{figure*}[t]
    \centering
    \begin{subfigure}[b]{0.70\textwidth}
        \includegraphics[width=\columnwidth]{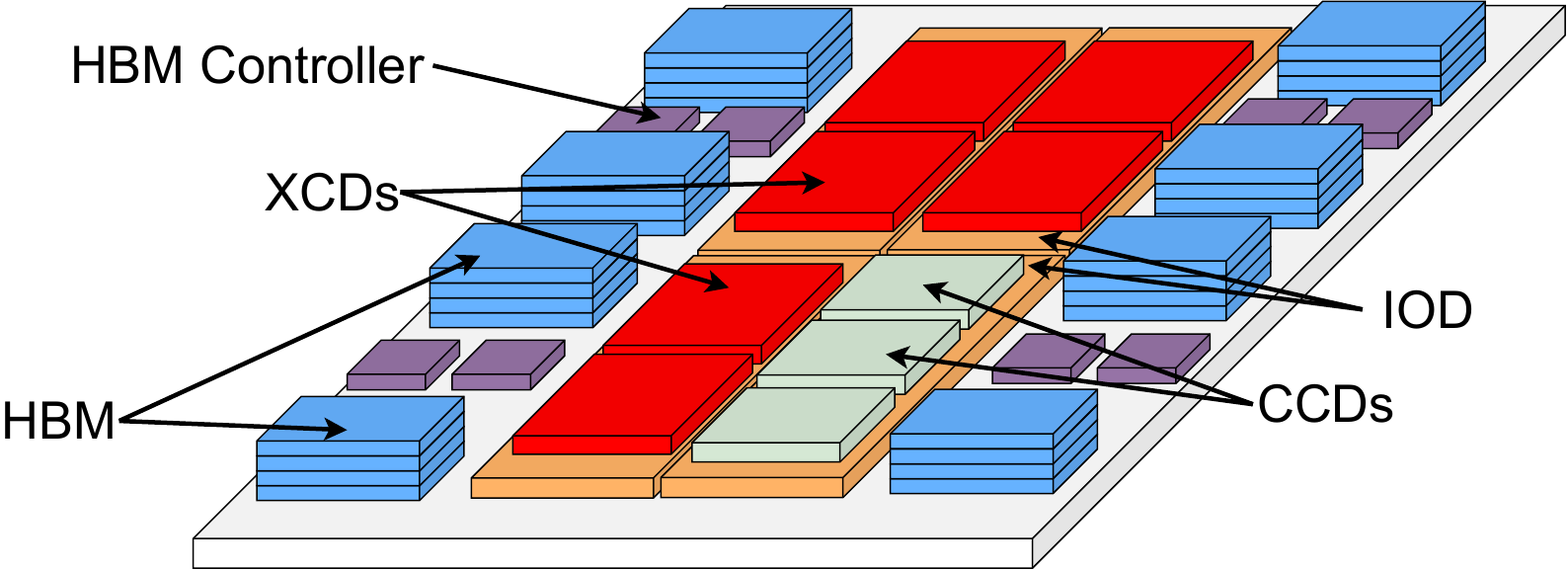}
        \caption{}
    \end{subfigure}
    
    \begin{subfigure}[b]{0.32\textwidth}
        \includegraphics[width=\columnwidth]{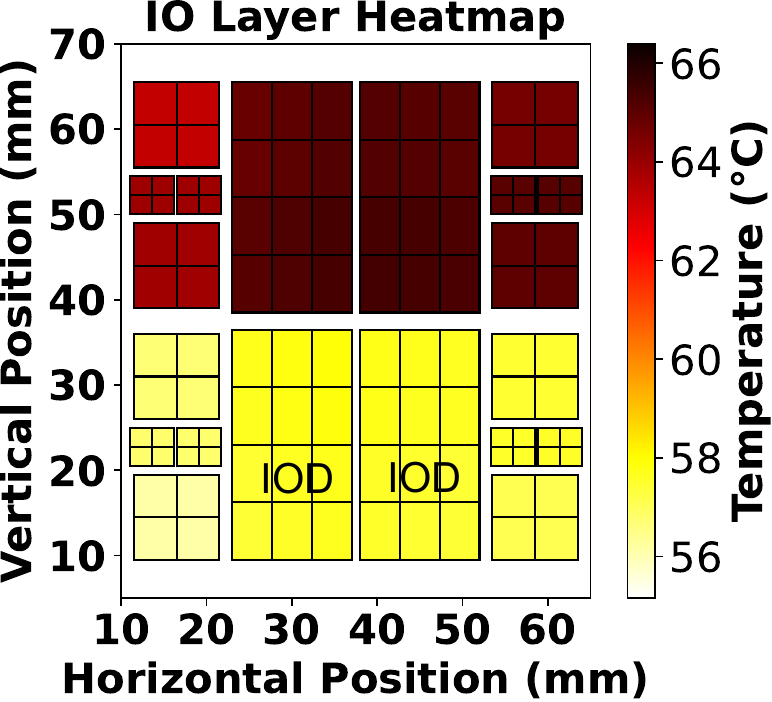}
        \caption{}
    \end{subfigure}
    \hfill
    \begin{subfigure}[b]{0.32\textwidth}
        \includegraphics[width=\linewidth]{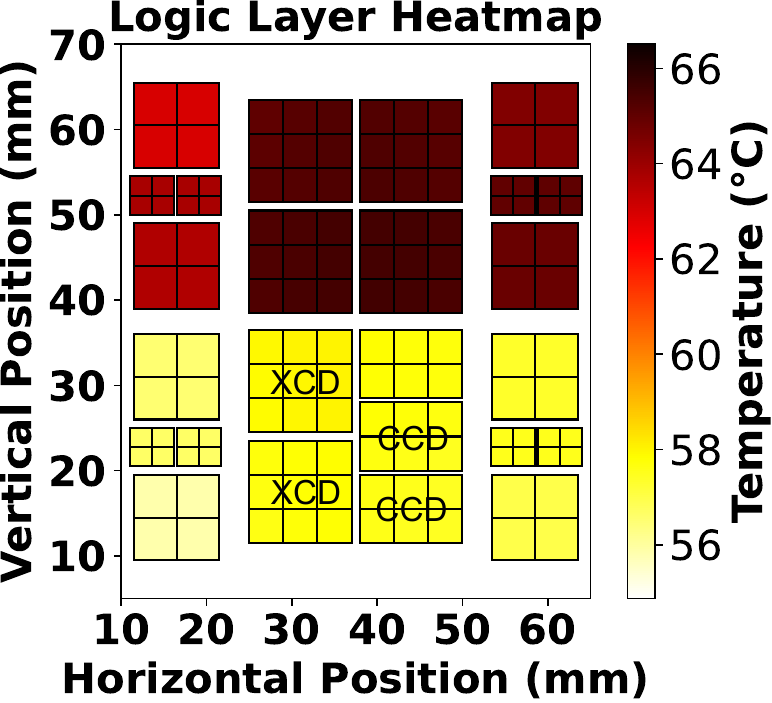}
        \caption{}
    \end{subfigure}
    \hfill
    \begin{subfigure}[b]{0.32\textwidth}
        \includegraphics[width=\linewidth]{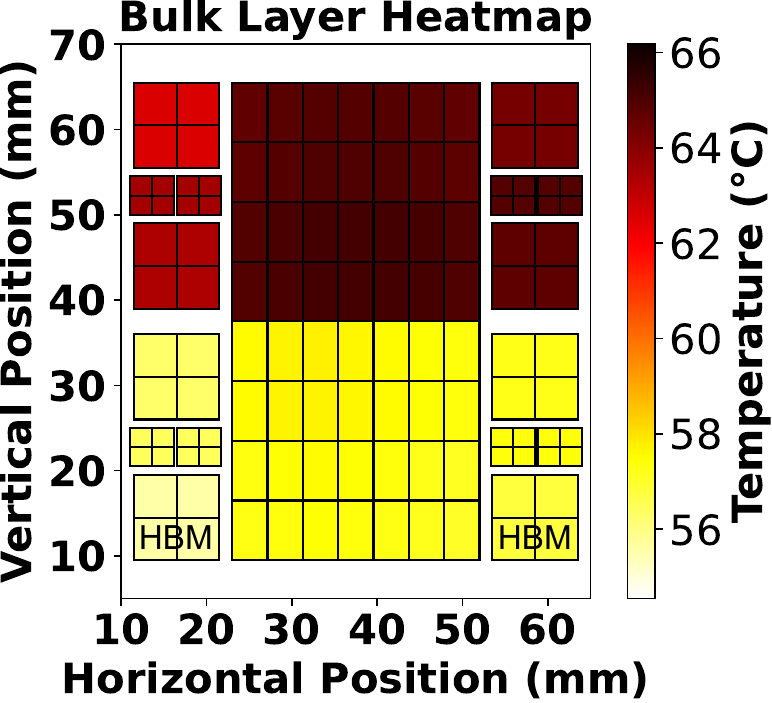}
        \caption{}
    \end{subfigure}
    \caption{\rev{Thermal modeling of AMD MI300A, a CPU-XPU heterogeneous system (a) Block diagram of MI300A architecture~\cite{smith2024realizing}, (b) Heatmap of IO Die layer, (c) Heatmap of compute die layers, (d) Heatmap of bulk-si layer.}}
    \label{fig:mi300}
\end{figure*}

\rev{To demonstrate the applicability of MFIT to complex heterogeneous architectures, we model AMD’s MI300A system in this section. 
The MI300A features a hybrid 2.5D/3D integration, where multiple chiplets are stacked across two tiers~\cite{smith2024realizing}. 
Specifically, IO dies (IODs) are placed on the bottom tier, with a mix of accelerator complex dies (XCDs) and CPU complex dies (CCDs) stacked directly above them. Our modeled configuration includes 6 XCDs and 3 CCDs. 
This compute structure is surrounded by 8 high-bandwidth memory (HBM) stacks and their corresponding controllers, as illustrated in Fig.~\ref{fig:mi300}(a). 
We model this system using our thermal RC framework with a total of 500 nodes. As in Section~\ref{ssec:setup}, we assign higher node density to active components such as the XCDs and CCDs, and lower node density to passive components like the lid and substrate. This demonstrates \textit{MFIT’s capability to model complex 2.5D/3D structures with heterogeneous node densities}.}

\rev{Conventional thermal simulators such as HotSpot, PACT, and 3D-ICE are not capable of accurately modeling such architectures. 
These tools typically assume homogeneous material layers and cannot model anisotropic materials or bidirectional heat dissipation, these capabilities that are critical for capturing the thermal behavior of high-performance industrial systems like MI300A. 
Fig.~\ref{fig:mi300}(b-d) shows temperature heatmaps across three vertical stacks, with labeled components, highlighting the spatial thermal variations across the heterogeneous chiplet layout.
}

% This section is short but I dont think it needs much
\begin{comment}
\bh{Accuracy comparison to HotSpot~\cite{skadron_temperatureAware}:} 
Table~\ref{tab:combined_accuracy} also lists the MAE of HotSpot simulations compared to FEM simulations for 2.5D and the 3D system. 
For the 2.5D system results, the average error of HotSpot is 0.95, 0.05, and 0.35 degrees greater than our thermal RC and DSS models for the 16, 36, and 64 chiplet systems, respectively. 
Notably, HotSpot also has lower accuracy in detecting temperature violations, but our primary advantage is in execution time, as discussed in Section~\ref{ssec:exe_time}.
For the 3D results, HotSpot's average error is 0.22 degrees greater than that of our thermal RC and DSS models. HotSpot again shows lower accuracy in detecting temperature violations in workloads 1, 4, and 6.
The larger error across system configurations can be attributed to the tuning of thermal RC model parameters based on the available reference FEM thermal model results. Additionally, HotSpot does not include the ability to provide different thermal conductivities along the x, y, and z-axis, further decreasing accuracy. 
\end{comment}

\section{Conclusion} \label{sec:conclusion}

% \todo{\textbf{Update this section once all other writing is complete}}

Due to increasing manufacturing costs, conventional monolithic 2D chips cannot sustain the increasing performance and compute capacity demands. 2.5D and 3D chipset systems have emerged as cost-effective solutions to continue the required scaling.
However, substantial compute power in a small volume intensifies the power density, leading to severe heat dissipation and thermal challenges. 
There is a strong need for open-source thermal modeling tools that enable researchers to analyze thermal behavior and perform thermally aware optimizations. Re-purposing existing approaches developed for monolithic chips incurs accuracy and execution time penalties, while custom-designed singular solutions have limited scope. 
To fill this gap, this paper proposed MFIT, a set of multi-fidelity thermal models that span a wide range of accuracy and execution time trade-offs. Since the proposed models are consistent by construction, designers can use them throughout the design cycle, from system specification to design space exploration and runtime resource management.

\bibliographystyle{ACM-Reference-Format}
\bibliography{reference/spec, reference/related,reference/ogras_paper, reference/iarch}

\end{document}